\documentclass[twocolumn, trackchanges]{aastex701}

\usepackage[version=4]{mhchem}
\usepackage{booktabs}
\usepackage{graphicx}
\usepackage{subcaption}
\usepackage{placeins}
\usepackage{hyperref}
\usepackage{tikz}
\usepackage{tikzpagenodes}
\usetikzlibrary{calc}

\shorttitle{Observation of spontaneous N-bearing PAH formation using ion trap}
\shortauthors{Siddhartha S. Payra et al.}

\begin{document}

\title{Observation of spontaneous N-bearing PAH formation using ion trap: A new formation pathway in the interstellar medium}

\author[orcid=0009-0007-7449-6108]{Siddhartha S. Payra}
\affiliation{Department of Physics, Indian Institute of Technology Madras, Chennai, India}
\email{ph23d029@smail.iitm.ac.in}  

\author[orcid=0009-0007-1425-6620]{Pratikkumar Thakkar}
\affiliation{Department of Physics, Indian Institute of Technology Madras, Chennai, India}
\email{ph23d007@smail.iitm.ac.in}

\author{Shiv Gupta}
\affiliation{Department of Chemistry, Indian Institute of Technology Madras, Chennai, India}
\email{shivg9097@gmail.com}

\author{Ruth Ann Mathews}
\affiliation{Department of Chemistry, Indian Institute of Technology Madras, Chennai, India}
\email{cy24d019@smail.iitm.ac.in }

\author[orcid=0009-0001-7766-8848]{Yash Lenka}
\affiliation{Department of Physics, Indian Institute of Technology Madras, Chennai, India}
\email{ph22d200@smail.iitm.ac.in}

\author[orcid=0000-0002-7329-7760]{Saurav Dutta}
\affiliation{Department of Physics, Indian Institute of Technology Madras, Chennai, India}
\email{sdutta@physics.iitm.ac.in}

\author[orcid=0009-0003-5527-0761]{Nihar Ranjan Behera}
\affiliation{Department of Physics, Indian Institute of Technology Madras, Chennai, India}
\email{nihar.ranjan.behera@physics.gu.se}

\author[orcid=0000-0002-1760-2898]{Krishna R. Nandipati}
\affiliation{Department of Chemistry, Indian Institute of Technology Madras, Chennai, India}
\email{knandipati@iitm.ac.in}

\author[orcid=0000-0003-4648-7136]{G. Aravind}
\affiliation{Department of Physics, Indian Institute of Technology Madras, Chennai, India}
\email[show]{garavind@iitm.ac.in}

\begin{abstract}
Nitrogen-bearing polycyclic aromatic hydrocarbons (N-PAHs) are key precursors to complex organic molecules in both the interstellar medium and the nitrogen-rich planetary atmospheres. Despite the recent detections of nitrogen-functionalized astromolecules, their formation pathways remain an open question.
The discrepancies between their predicted and observed abundances point to unknown mechanisms such as ion-molecule reactions that govern their evolution in the astrophysical environments.
Employing an ion trap technique in conjunction with electronic structure calculations, we unravel multiple barrier-less reactions between gas-phase pyrimidine cations (C$_4$H$_4$N$_2^+$) and acetylene (C$_2$H$_2$) which form a hitherto unreported endocyclic- N-PAH (C$_8$H$_7$N$_2^+$) at room temperature. The present measurements on reactions involving a double-nitrogen substituted aromatic heterocycle have implications to the astrochemistry of both the Titan’s atmosphere and the interstellar medium.
\end{abstract}



\section{Introduction} 
Aromatic molecules in space have profoundly influenced the chemical evolution of the universe.
In particular, polycyclic aromatic hydrocarbons (PAHs) account for approximately 15\% of interstellar carbon and contribute about 20\% of the cosmos's infrared emission~\citep{li2020spitzer,allamandola2021pah}.
The series of unidentified infrared bands between 3 and 20 $\mu$m, most notably at 3.3, 6.2, 7.7, 8.6, 11.3, and 12.7 $\mu$m arise from the vibrational transitions of both neutral and ionized PAH species~\citep{li2020spitzer}.
A slight shift was observed in the $6.2$ $\mu$m emission band's peak position, which has been proposed as a potential indicator of nitrogen-bearing polycyclic aromatic hydrocarbons (N-PAHs) existence within the PAHs population that produce this emission~\citep{mattioda2008near,peeters2002rich,hudgins2005variations,canelo2018variations}.
More recently, rotational spectroscopy at radio frequency has unambiguously detected nitrogen-functionalized coronene~\citep{wenzel2025discovery}, pyrene~\citep{wenzel2024detection,wenzel2025detections}, and naphthalene~\citep{mcguire2021detection} in dense molecular clouds.
Furthermore, in situ measurements from NASA's Cassini mission have revealed a rich population of N-PAHs in the atmosphere of Titan, highlighting the prevalence of nitrogen-bearing aromatics in planetary environments~\citep{vuitton2007ion,ali2015organic,crary2009heavy,haythornthwaite2021heavy,waite2005ion}.

\par The formation mechanisms of PAHs and N-PAHs throughout the space remain an open question, and astronomical observations alone are insufficient to fully elucidate these mechanisms.
The ambiguity between predicted and observed molecular abundances highlights the necessity for experimental studies that investigate astrochemical reaction networks~\citep{payra2025dissociative, payra2026wavelength}, including key reaction pathways and their associated rate coefficients.
This requires the study of neutral-neutral, radical-neutral, radical-radical, and ion-neutral reactions~\citep{kaiser2021aromatic,sandford2020prebiotic,doddipatla2021low,rap2022low,rap2024noncovalent, yang2024low}.
Neutral-neutral molecular reactions typically require substantial activation energy, making them less probable to take place~\citep{sandford2020prebiotic}.
In contrast, ion-molecule reactions often proceed without any activation energy, rendering them more favorable.
Although dense molecular clouds lack direct stellar radiation, they are not entirely free from ionizing influences.
Substances within these clouds are subjected to ionizing processes driven by cosmic rays, and secondary electrons produced through the cosmic ray ionizations.
These energetic electrons collisionally excite molecular hydrogen, which subsequently emits ultraviolet (UV) photons upon radiative de-excitation in a process known as the Prasad-Tarafdar mechanism~\citep{prasad1983uv,sandford2020prebiotic}.
This internally generated UV field sustains ionization and drives further chemical complexity despite the radiation shielding, rendering cosmic-ray-induced ionization and resulting cascades of ion-neutral reactions as crucial components in the astrochemical synthesis of PAHs and N-PAHs.
In addition to the cold dense molecular clouds, gas-phase ion-molecule reactions can occur in high-temperature regions like protostellar disks and circumstellar envelopes, where regions of higher density and temperature can accelerate these processes.
To investigate the growth of N-PAHs through ion-neutral reactions, several studies over the past decade have investigated reactions between aromatic ions, such as pyridine (\ce{C5H5N^{+}}), aniline(\ce{C6H5NH2^{+}}), and benzonitrile (\ce{C7H5N^{+}}) with acetylene (\ce{C2H2})~\citep{soliman2013formation,rap2022low,rap2024noncovalent,shiels2021reactivity}.
\par However, experimental data obtained under astrophysically relevant conditions are lacking, preventing the incorporation of validated reaction mechanisms and \textcolor{black}{rate coefficients} into models for accurate prediction of molecular abundances.
Notably, prior investigations of ion-molecule reactions with astrochemical relevance have predominantly focused on systems featuring single nitrogen substitutions or cyano-functionalized aromatic rings.
To elucidate the impact of double nitrogen substitution on molecular growth processes, we selected pyrimidine(\ce{C4H4N2}), a key precursor to nucleobases and already discovered in carbonaceous chondrites, including Murchison. 
This choice is further motivated by data from Cassini’s Ion and Neutral Mass Spectrometer (INMS), which detected a signal at mass 81 that may correspond to protonated pyrimidine in Titan’s atmosphere~\citep{nixon2024composition,vuitton2007ion}.
\par Here, we present an ion–neutral kinetic study of the reaction between pyrimidine cations (\ce{C4H4N2^+}) and acetylene.
A previous study of this reaction was conducted at elevated pressures and high collision energies~\citep{soliman2013formation}, conditions that are not representative of those in astronomical environments.
To better assess the astrochemical significance of this process, we performed kinetic measurements at reduced pressures, utilizing a 22-pole radio frequency ion trap apparatus.
This approach enables the investigation of sequential ion–molecule reactions under conditions that closely mimic those found in space, thereby providing critical experimental constraints for astrochemical models.
Kinetic rate coefficient measurements demonstrated an efficient radiative association during the initial addition of acetylene, followed by a subsequent exothermic bimolecular reaction that yields the stable nitrogen-bearing bicyclic cation product. In order to corroborate these experimental observations, we have also performed electronic structure calculations to infer the structure of the intermediates (INT)/transition states (TS) and product (P) ions that help identify the plausible reaction pathway leading to the product formation.
The calculations confirm that the reaction is exothermic, with the computed optimized structures of INT/TS lying below the entrance channel of the reactant.

\section{Methods}

\textbf{Experimental kinetic studies.} All experiments were performed using a modified 22-pole radio frequency ion trap setup. The schematic is provided in Appendix A.
A detailed description of the instrument can be found in ref.~\citep{behera202422}; only a brief description of the experimental setup and the procedure is given here.
Pyrimidine (purity 98\%, Sigma-Aldrich) was vaporized by heating its reservoir and subsequently introduced into the interaction chamber using a pulsed solenoid valve.
High-purity helium (99.999\%) at 1 atmospheric pressure served as the carrier gas.
Electrons with energies ranging from 100 to 105 eV were directed onto the pyrimidine molecular beam at 150 mm downstream of the nozzle.
The cations formed after the interaction were repelled toward the first quadrupole mass spectrometer (QMS-1).
An Einzel lens mounted immediately after QMS-1 focused the mass-selected ions into the 22-pole ion trap.
A short pulse of the helium buffer gas was introduced into the ion trap using a pulsed solenoid valve to cool the ions kinetically.
The neutral reaction partner, acetylene (99.6\% pure acetylene dissolved in acetone), was continuously injected into the trap through a needle valve to initiate the ion-molecule reaction for defined trapping times ranging from 0 to 1000 ms.
The contents of the trap were analyzed as a function of storage time using a second quadrupole mass spectrometer (QMS-2, \textcolor{black}{with a mass resolution (M/$\Delta$M) of 2000}) and a Channel Electron Multiplier (CEM) detector.
Multiple measurements were taken at each condition to estimate the experimental error.
The number density of the neutral gas was determined by measuring the pressure using a Pfeiffer CMR 375 capacitive gauge, which was directly connected to the trap volume, and offers an accuracy of 0.15\% within its operating pressure range of 0.001 Pa to 11 Pa.
Some of our experiments require them to be conducted at a neutral gas pressure lower than the working range of the capacitive gauge.
Hence, we used two additional gauges to measure the pressure of the 22-pole trap chamber and the QMS-1 chamber, utilizing a Pfeiffer Compact Full Range Gauge and a Compact Cold Cathode Gauge, with an accuracy of 30\% and a lower cutoff pressure of 10$^{-7}$ Pa, respectively.
To calibrate these gauges against the capacitive gauge, high densities of buffer gas were established inside the 22-pole trap and varied linearly to compute a calibration function which can be extrapolated to the low density regime.
The typical acetylene number density between $1\times 10^{11}$ cm$^{-3}$ and $4\times 10^{13}$ cm$^{-3}$ was used.
The kinetic profiles were analyzed by fitting them to a set of ordinary differential equations (ODE) that model different reaction mechanisms, and first order rate coefficients were extracted as described in Appendix B.
The second order rate coefficients for the individual reactions were obtained from linear fits of the first order \textcolor{black}{rate coefficients} measured at different acetylene number densities.

\par \textbf{Electronic structure calculations.} The electronic structure calculations were performed at the level of Density Functional Theory (DFT)/B3LYP using the Gaussian 16 software package~\citep{g16}. For all calculations, the 6-311G(d) basis set was used to balance accuracy and computational cost. Moreover, tight convergence criteria and a superfine integration grid were employed during the optimizations of reactant and product geometries, as well as intermediate/transition state structures (see Appendix D \& E for the initial guesses and optimized coordinates for selected geometries). The frequencies of normal modes for all the optimized geometries were calculated -- the transition state geometries were characterized by a single imaginary frequency, whereas the reactant and product equilibrium geometries by all real frequencies. The computed relative electronic energies of the optimized structures, including zero-point vibrational energy (ZPE) correction, are mapped to construct a reaction pathway profile.

\section{Results}
\begin{figure}
	\centering
	\includegraphics[width=0.95\columnwidth]{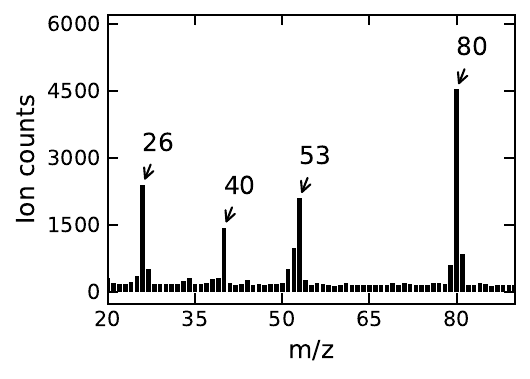} 
	\caption{Electron-impact mass spectrum of pyrimidine with 100 eV electrons. In addition to the parent ion peak at m/z 80, fragments corresponding to HCN loss were observed.
		}
	\label{fig1}
\end{figure}

\textbf{Ion–molecule reaction kinetics.} Figure~\ref{fig1} shows the mass spectrum obtained following the electron-impact ionization of the pyrimidine/helium mixture. The parent cation at m/z 80 is observed with a dominant yield. The cation peak at m/z 53 corresponds to HCN loss, while the peaks at m/z 40 and 26 correspond to \ce{C4H4N2^{2+}} and \ce{C2H2^+}, respectively.
\par
\begin{figure} 
	\centering
	\includegraphics[width=0.95\columnwidth]{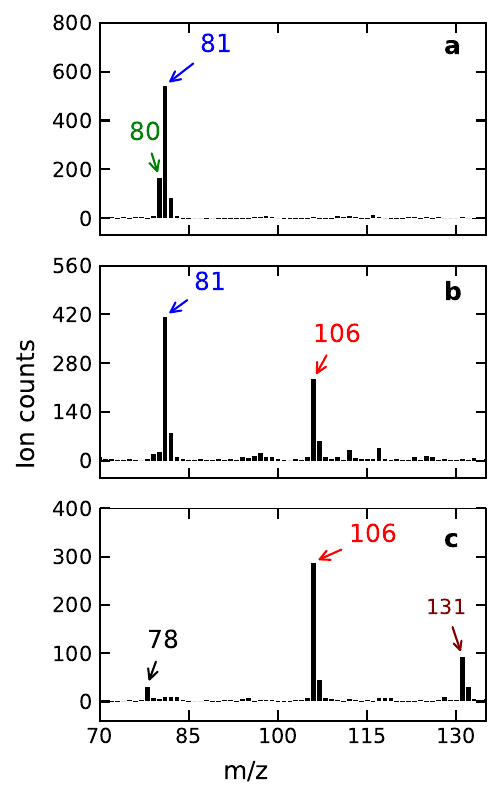} 
	\caption{Product mass spectra from ion-molecule reactions of pyrimidine ions with acetylene. Mass spectra showing the reaction products formed from pyrimidine (\ce{C4H4N2^{+}}) and protonated pyrimidine (\ce{C4H5N2^{+}}) ions reacting with acetylene (\ce{C2H2}) under different conditions: (\textbf{a}) control spectrum with no acetylene present, (\textbf{b}) low acetylene number density ((2.4$\pm$0.8) $\times$ 10$^{11}$ cm$^{-3}$), and (\textbf{c}) high acetylene number density ((2.7$\pm$0.9) $\times$ 10$^{13}$ cm$^{-3}$). Formation of new product species was observed in the presence of acetylene.}
	\label{fig2}
\end{figure}

 The pyrimidine parent cations (\ce{C4H4N2^{+}}, m/z 80) were mass-selected using a quadrupole mass spectrometer (QMS-1) and loaded into the 22-pole ion trap. After a specified storage time, the trapped ions were extracted and mass-analyzed using QMS-2 to obtain the spectrum shown in Fig.~\ref{fig2},(\textbf{a}).
Following the trapping with helium buffer gas at 295 K, protonation was consistently observed to yield the cation at m/z 81. This protonation is attributed to residual proton donors such as the residual moisture in the helium buffer gas.
The nitrogen atoms in pyrimidine exhibit a high proton affinity of $\approx 879$ kJ mol$^{-1}$~\citep{nguyen1997protonation}, making them preferential protonation sites even with trace sources such as background moisture~\citep{kumar2019effects}. 

\begin{figure*}[ht!]
	\centering
	\includegraphics[width=0.95\textwidth]{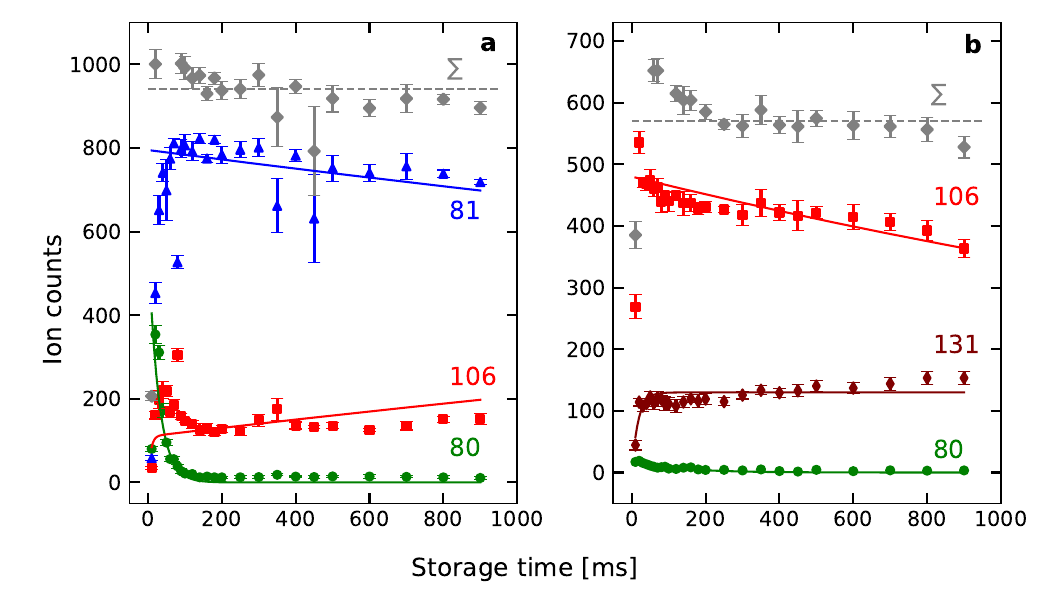}
	\caption{Kinetic profiles of reactants and products in ion-molecule reactions between pyrimidine and acetylene. Time-resolved ion counts show the reactions of pyrimidine (\ce{C4H4N2^{+}}, m/z 80) and protonated pyrimidine (\ce{C4H5N2^+}, m/z 81) with acetylene (\ce{C2H2}, m/z 26) at 295 K under two reaction conditions: (\textbf{a}) low acetylene number density ((2.4$\pm$0.8) $\times$ 10$^{11}$ cm$^{-3}$) and (\textbf{b}) high number density ((2.7$\pm$0.9) $\times$ 10$^{13}$ cm$^{-3}$). Data points represent experimental ion counts as a function of storage time, with error bars indicating measurement uncertainties. Solid lines represent kinetic fits obtained using an ODE model. The grey dots indicate the sum of all ions ($\sum$) in the ion trap. Initial data points \textcolor{black}{up to 20 ms} were excluded from the analysis.}
	\label{fig3}
\end{figure*}

\begin{figure*}
	\centering
	\includegraphics[width=0.9\textwidth]{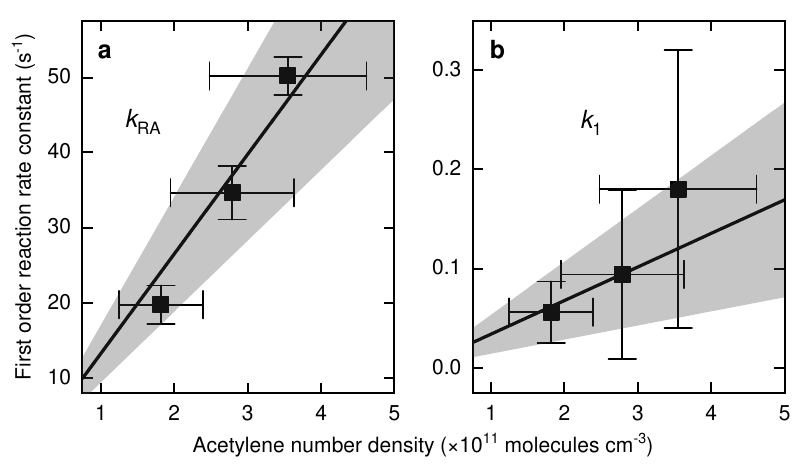}
	\caption{First order rate coefficients derived from the kinetic profiles of (\textbf{a}) m/z 80 and (\textbf{b}) m/z 81 as a function of the acetylene number density. The second order rate coefficients for the reactions of m/z 80 (\textit{k}$\mathrm{_{RA}}$) and 81 (\textit{k}$\mathrm{_1}$) with acetylene were determined from linear least-squares fits to the data. Error bars represent the 1$\sigma$ uncertainties, while the grey shaded regions denote the 95\% confidence bands of the fits.}
	\label{second_order_rate}
\end{figure*}

\par Ion-molecule reactions between the parent and protonated pyrimidine cations with acetylene were studied by letting in acetylene gas (m/z 26) into the trap when these two cations were stored.  
The \ce{C2H2} molecules reacted rapidly with the pyrimidine ions and formed new molecules of heavier mass, as shown in the Fig.~\ref{fig2},(\textbf{b,c}). At lower number densities of acetylene, (2.4$\pm$0.8) $\times$ 10$^{11}$ cm$^{-3}$, we observed the formation of m/z 106 and 107 (Fig.~\ref{fig2},(\textbf{b})) and subsequently the formation of m/z 131 at a higher number density of (2.7$\pm$0.9) $\times$ 10$^{13}$ cm$^{-3}$, as shown in Fig.~\ref{fig2},(\textbf{c})).
The yield of the cations was measured for various ion-storage times. 
These measurements were carried out several times with precisely measured acetylene number densities, and two representative kinetic profiles are presented in Fig.~\ref{fig3}.

\par At low density of acetylene number ((2.4$\pm$0.8) $\times$ 10$^{11}$ cm$^{-3}$), we observe a decrease in the number of parent and protonated cations with time. Almost all of the stored parent cations at m/z 80 were observed to react and form cations of m/z 106 within 200 ms.
In addition, the increase of m/z 106 at later times indicates the formation of m/z 106 from m/z 81 (Fig.~\ref{fig3},(\textbf{a})).
At room temperature and low acetylene densities, the formation of m/z 106 (\ce{C4H4N2^{+}}) occurs via an acetylene association process, where stabilization requires emission of a photon (radiative association) or collision with a third body (termolecular association).
Typically, at low number densities of acetylene ($\sim10^{9}-10^{11}$ cm$^{-3}$), the reactions proceed via radiative association~\citep{rap2024noncovalent,rap2022low}.
To further investigate the reaction, the first order reaction \textcolor{black}{rate coefficients} calculated from the fitting of the ODE model to the reaction kinetic profiles were plotted as a function of the acetylene number density, as shown in Fig.~\ref{second_order_rate},(\textbf{a}). Within the density range explored, the first order \textcolor{black}{rate coefficients} exhibit a linear dependence on the acetylene number density, consistent with radiative association kinetics. The \textcolor{black}{rate coefficient} obtained from the slope of the linear fit can thus be regarded as an upper limit to the radiative-association \textcolor{black}{rate coefficient}.
The experimentally determined rate coefficient for the radiative association to form the m/z 106 cation is \textit{k}$\mathrm{_{RA}}=(1.3\pm0.3)$ $\times$ 10$^{-10}$ cm$^3$ s$^{-1}$.
Similar radiative association rate coefficients have been reported for reactions involving other large aromatic molecules.
For instance, pyridine reacting with acetylene shows values around 3 $\times$ 10$^{-10}$ cm$^3$ s$^{-1}$~\citep{shiels2021reactivity} and (8.0$\pm$3.5) $\times$ 10$^{-11}$ cm$^3$ s$^{-1}$~\citep{rap2022low}).
Likewise, benzonitrile combined with acetylene exhibits \textcolor{black}{rate coefficients} of (4.2$\pm$2.5) $\times$ 10$^{-11}$ cm$^3$ s$^{-1}$ ~\citep{soliman2015growth} and (3.8$\pm$0.4) $\times$ 10$^{-11}$ cm$^3$ s$^{-1}$~\citep{rap2024noncovalent}.
A comparable, relatively high \textcolor{black}{rate coefficient} ((1.4$\pm$1.2) $\times$ 10$^{-9}$ cm$^3$ s$^{-1}$~\citep{soliman2013formation}) was previously reported for the ion-molecule reaction between pyrimidine and acetylene at 308 K using ion mobility tandem mass spectrometry. 
This high reaction \textcolor{black}{rate coefficient} is plausibly due to the influence of termolecular stabilization, which resulted from the higher pressures ($\sim7$ Pa or higher) employed in that experiment. 
In this work, we present the radiative association rate coefficient, while the previous tandem mass spectrometry work~\citep{soliman2013formation} likely yielded a saturated termolecular rate coefficient. The termolecular \textcolor{black}{rate coefficients} can be up to an order of magnitude greater than that for the radiative association, as previously observed by Rap et al.~\citep{rap2022low} in their study of pyridine and acetylene reaction.

\par In addition to the radiative association between the parent cation and \ce{C2H2}, a reactive two-body collision process involving hydrogen-abstraction and acetylene-addition (HACA) can form m/z 106 from the protonated pyrimidine (\ce{C4H5N2^{+}}).
The measured reaction \textcolor{black}{rate coefficient} for this bimolecular reaction is \textit{k}$\mathrm{_1}=(3.4\pm1.1)$ $\times$ 10$^{-13}$ cm$^3$ s$^{-1}$ (see Fig.~\ref{second_order_rate},(\textbf{b})).
A small abundance of m/z 107 is also observed, which can be formed through a two-body association process of \ce{C4H5N2^{+}} and acetylene.

\begin{table*}[ht!]
	\centering
	\caption{Experimentally determined reaction rate coefficients. The various reaction pathways and rate coefficients of the ion–molecule reaction between pyrimidine (\ce{C4H4N2^+}), protonated-pyrimidine (\ce{C4H5N2^+}), and acetylene (\ce{C2H2}) at 295 K.}
	\label{tab:1} 
	
	\begin{tabular}{lcccc}
		\\
		\hline
        \hline
		Reaction & Rate coefficient & Type & Label & Number density of \ce{C2H2}\\
		 & (cm$^3$ s$^{-1}$) & & & (cm$^{-3}$)\\
         \hline
		\ce{80^+ -> 106^+} & (1.3$\pm$0.3) $\times$ 10$^{-10}$ & Radiative association & \textit{k}$\mathrm{_{RA}}$ & $1.3        \times 10^{11}-3.9 \times 10^{11}$ \\
            \ce{81^+ -> 106^+} & (3.4$\pm$1.1) $\times$ 10$^{-13}$ & Bimolecular & \textit{k}$\mathrm{_1}$ & $1.3 \times 10^{11}-3.9 \times 10^{11}$ \\
            \ce{106^+ -> 131^+} & (2.2$\pm$0.9) $\times$ 10$^{-12}$ & Bimolecular & \textit{k}$\mathrm{_2}$ & $(2.7\pm0.9)\times 10^{13}$ \\
		\hline
	\end{tabular}
    \tablecomments{The uncertainties represent one standard deviation (1$\sigma$).}
\end{table*}

\par At higher acetylene concentrations, essentially all the trapped parent and protonated parent cations were observed to undergo reaction and form the 106 cation.
Further, as shown in Fig.~\ref{fig3},(\textbf{b}), the m/z 106 cations further reacted with \ce{C2H2} and a dehydrogenated product at m/z 131 is observed.
The sequential reaction leading to the formation of the m/z 131 cation proceeds via a bimolecular reaction with a rate coefficient of \textit{k}$\mathrm{_2}=(2.2\pm0.9)$ $\times$ 10$^{-12}$ cm$^3$ s$^{-1}$.
\textcolor{black}{The relatively smaller increase in the m/z 131 compared with the decrease in the m/z 106 is likely due to the lower trapping and extraction efficiencies of the heavier ion. Therefore, the rate coefficient, \textit{k}$_2$ was determined from the depletion of the m/z 106 ion, as described in Appendix B.}
At this high concentration regime, the ion with m/z 132 (Fig.~\ref{fig2},(\textbf{c})) is likely formed via a termolecular association process, wherein the m/z 106 ion associates with \ce{C2H2} and is subsequently stabilized through collisional interaction with a third body.
A summary of all the experimentally determined \textcolor{black}{rate coefficients} is provided in Table \ref{tab:1}.

\begin{figure*}[ht!]
	\centering
	\includegraphics[width=0.95\textwidth]{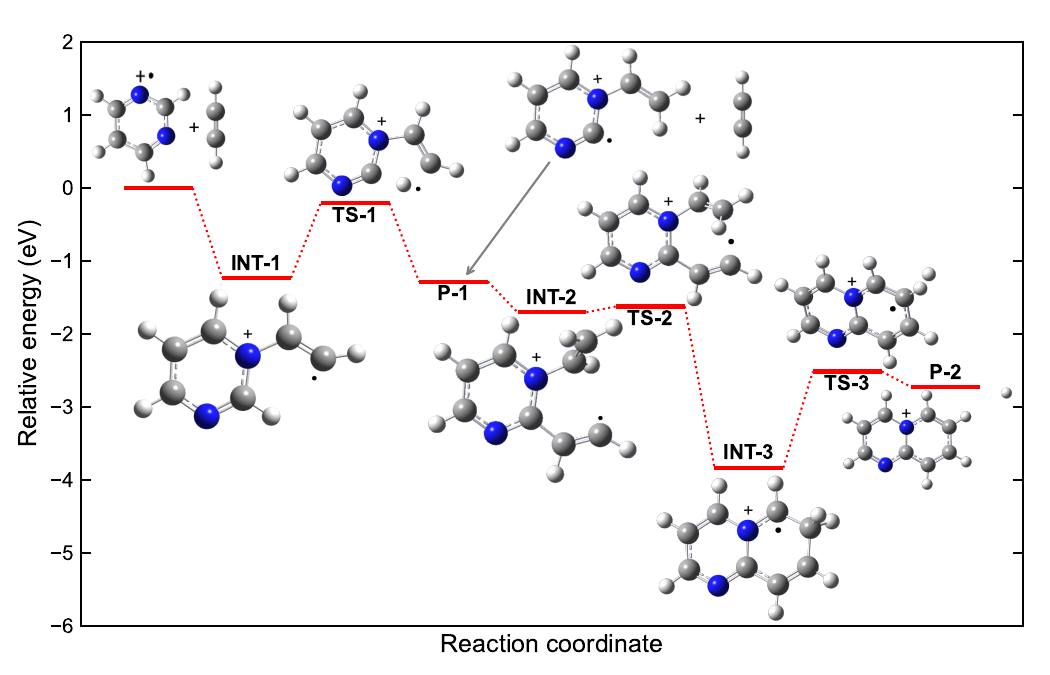} 
	\caption{Calculated electronic energies of the various structures involved in the reaction of pyrimidine$^+$ (\ce{C4H4N2^+}) with acetylene (\ce{C2H2}) to form the bicyclic aromatic product \ce{C8H7N2^+} (P-2) with m/z 131. The energy profile begins with \ce{C4H4N2^+} and \ce{C2H2} at infinite separation on the left of the abscissa and terminates upon formation of \ce{C8H7N2^+} on the right. The electronic energies are calculated at the B3LYP/6-311G(d) level of theory.
		}
	\label{fig4}
\end{figure*}

\par
A minor ion signal corresponding to m/z 78 is detected, at higher acetylene densities ((2.7$\pm$0.9) $\times$ 10$^{13}$ cm$^{-3}$), as illustrated in \ref{fig3}(\textbf{c}).
This ion is tentatively assigned to \ce{C6H6^+}, which may form via a dissociative charge transfer mechanism involving the \ce{C4H4N2^{+}(C2H2)_{3}} complex.
Given that benzene possesses a slightly lower ionization potential (IP = 9.2 eV) compared to pyrimidine (IP = 9.3 eV), partial charge transfer from the pyrimidine cation to the acetylene cluster can trigger cyclization of three acetylene molecules, resulting in the formation of a benzene (\ce{C6H6}) unit~\citep{soliman2012formation, momoh2008formation}.
Subsequently, this benzene moiety can fully abstract the charge from the pyrimidine ion, producing \ce{C6H6^+} through the ejection of neutral pyrimidine.
A similar phenomenon has been previously reported by Soliman et al.~\citep{soliman2013formation} at a very high pressure of acetylene (1.3 Torr).

\par There are a few weaker mass peaks (m/z 112, 117, etc.), which are likely due to the reactions of the primary ion with the residual gases in the trap.
A trace amount of acetone is present in the acetylene cylinder. Although no signal is observed at m/z 58, very weak peaks at m/z 138 and 139 are detected, as shown in the wide-range mass spectrum in Appendix C. These ions can be assigned to the reactions of pyrimidine and protonated pyrimidine with acetone. Due to their negligible abundance and the absence of any measurable temporal evolution, these species were excluded from the kinetic analysis.

\par \textbf{Electronic structure calculations and reaction mechanisms.} Although direct experimental characterization of the ion structure is not feasible for us, we utilized quantum chemical calculations to determine the structure and energies of the products and TS/INT.
\textcolor{black}{An isomer search was first carried out for the product ion C$_8$H$_7$N$_2^+$ (m/z 131), and the relative energies of the identified isomers are presented in the Appendix F. Based on the lowest-energy isomer, a plausible reaction mechanism was proposed, as illustrated in Fig.~\ref{fig4}.}
The reaction mechanism unfolds through several consecutive steps that collectively describe the formation of the bicyclic aromatic product.
At first, the addition of acetylene to the nitrogen-radical cationic site of pyrimidine forms intermediate \textbf{INT-1}.
This addition is followed by a critical transfer of a hydrogen atom from the aromatic ring to the acetylene chain through transition state \textbf{TS-1}, ultimately yielding the first detectable product (\textbf{P-1}) at m/z 106.
Subsequently, hydrogen migration creates a new radical site on the aromatic ring, which facilitates the attachment of a second acetylene molecule to this activated carbon center (\textbf{INT-2}). The \textbf{INT-2} undergoes ring closure through \textbf{TS-2} with a lowest energy barrier forming \textbf{INT-3}. This step is finally followed by an elimination of H-atom to form the final bicyclic product detected at m/z 131 \textbf{P-2}.
The mechanism governing the second acetylene addition closely parallels the well-established HACA pathway, demonstrating the fundamental similarity between these processes.
Two significant energy barriers are found, one for hydrogen migration from the ring to the side acetylene group (\textbf{INT-1}/\textbf{TS-1}) and the other for the hydrogen atom removal step (\textbf{INT-3}/\textbf{TS-3}).
However, both transition state energies (\textbf{TS-1} and \textbf{TS-3}) lie below the entrance energy.
Thus, the whole reaction provides an exothermic pathway (by 2.73 eV) for the formation of a bicyclic end product (\textbf{P-2}). The optimized TS geometries and the coordinates, and the frequency of a normal mode characterizing TS, i.e., with negative force constants, are presented in the supplementary text. In addition, the IRC calculations shown in Appendix G provide further confirmation of the reaction pathway suggested in Fig.~\ref{fig4}, which means that the TSs are smoothly connected to the corresponding reactant and products on the minimum energy path on a multidimensional potential energy surface. 

\begin{table*}[ht!]
	\centering
	\caption{Radiative association \textcolor{black}{rate coefficients} for various ion-molecule reactions relevant to astronomical environments and comparison with the determined reaction rate coefficient from this study.}
	\label{tab:2} 
	
	\begin{tabular}{cccc} 
		\\
		\hline
        \hline
		Reaction & Radiative association rate coefficient & Temperature & Ref.\\
		 & (cm$^3$ s$^{-1}$) & (K)&\\
		\hline
		\ce{C5H5^{+} +C2H2}&$(4.8\pm1.9)\times10^{-11}$&363&\cite{ozturk1989reactions}\\
        \ce{C6H5^{+} +C2H2}&$5.2-7.1\times10^{-10}$&300&\cite{soliman2012formation}\\
        \ce{C5H4NH^{+} +C2H2}&$3.0\times10^{-10}$&298&\cite{shiels2021reactivity}\\
        \ce{C5H5N^{+} + C2H2}&$(8.0\pm3.5)\times10^{-11}$& 150&\cite{rap2022low}\\
        \ce{C5H5N^{+} + C2H2}&$(9.0\pm5.0)\times10^{-11}$&304&\cite{soliman2013formation}\\
        \ce{C7H5N^{+} + C2H2}&$(3.8\pm0.4)\times10^{-11}$&150&\cite{rap2024noncovalent}\\
        \ce{C4H4N2^{+} + C2H2}&$(1.4\pm1.2)\times10^{-9}$&308&\cite{soliman2013formation}\\
        \ce{C4H4N2^{+} + C2H2}&(1.3$\pm$0.3)$\times$10$^{-10}$&295&This work\\
		\hline
	\end{tabular}
\end{table*}

\begin{figure*}[ht!]
	\centering
	\includegraphics[width=0.9\textwidth]{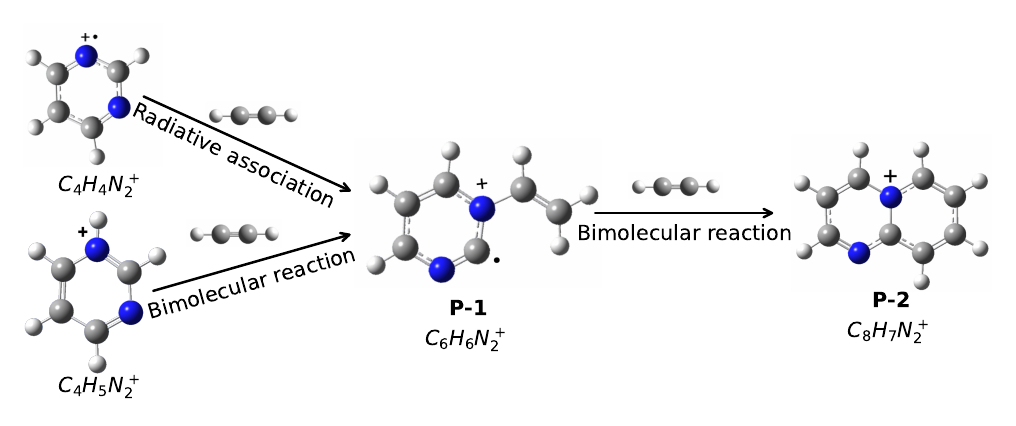} 
	\caption{Schematic overview of the overall reaction of pyrimidine$^+$ (\ce{C4H4N2^{+}}) and protonated-pyrimidine (\ce{C4H5N2^+}) with acetylene (\ce{C2H2}).
		}
	\label{fig5} 
\end{figure*}

\begin{figure}
	\centering
	\includegraphics[width=0.95\columnwidth]{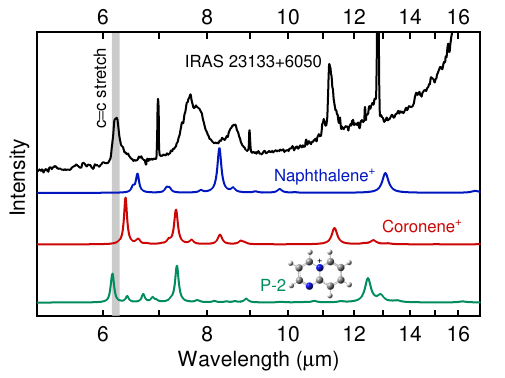} 
	\caption{Comparison of the observed interstellar infrared emission spectrum with the calculated infrared absorption spectra of the final reaction product \textbf{P-2} and representative cationic PAHs. The black curve shows the infrared emission spectrum of the compact H~\textsc{ii}-region Infrared Astronomical Satellite (IRAS) 23133+6050~\citep{sloan2003uniform}, with the astronomical 6.2 $\mu$m emission feature highlighted by the grey shaded region. The calculated absorption spectra of the final reaction product (green), together with those of naphthalene$^+$ (blue) and coronene$^+$ (red), are overlaid to compare the positions of the \ce{C=C} stretching vibrational bands (grey shading). The theoretical spectra of the final reaction product were calculated at the B3LYP/6-311G(d) level of theory.}
	\label{IR_comparision} 
\end{figure}

\section{Discussion and astrophysical implications}
Our experiments demonstrate the formation of a bicyclic nitrogen-bearing aromatic hydrocarbon via the ion-molecule reaction.
We showed how the incorporation of nitrogen into the aromatic ring enhances its reactivity, since there is no addition of acetylene with the benzene cation (\ce{C6H6^{+}}) at room temperature~\citep{momoh2008formation}.
Furthermore, our results indicate that the efficiency of the acetylene addition reaction increases even further when two nitrogen atoms are substituted into the aromatic ring, compared to a single nitrogen substitution ((9.0$\pm$5.0) $\times$ 10$^{-11}$ cm$^3$ s$^{-1}$ for pyridine~\citep{soliman2013formation} and (1.3$\pm$0.3) $\times$ 10$^{-10}$ cm$^3$ s$^{-1}$ for pyrimidine).
The derived radiative association rate coefficient is lower than the Langevin collision rate coefficient (2 $\times$ 10$^{-9}$ cm$^3$ s$^{-1}$). In contrast, \citet{soliman2013formation} reported a rate coefficient of (1.4$\pm$1.2) $\times$ 10$^{-9}$ cm$^3$ s$^{-1}$, which is comparable to the Langevin limit. The relatively high \textcolor{black}{rate coefficient} reported in the previous study may be attributed to the higher kinetic energy of the reactant ions and the elevated pressure under which the experiments were performed. In the present work, the lower collision energy and reduced pressure, together with the barrier associated with the H migration step \textbf{INT-1}/\textbf{TS-1} ($\sim$1.02 eV), are likely responsible for the slower radiative association rate coefficient observed.

\par The second acetylene addition involves the reaction of the \textbf{P-1} intermediate, which is formed through a preceding exothermic reaction ($\Delta H = -1.3$ eV) within the ion trap. The intermediate may retain a degree of internal excitation at the onset of the subsequent reaction. In the present work, the formation of \textbf{P-1} is proposed to proceed via radiative association, where the collision complex is stabilized through spontaneous photon emission. Therefore, part of the excess energy released during the association is dissipated radiatively, and the stabilized product is not expected to retain the full reaction exothermicity as internal excitation~\citep{jimenez2025ion, jusko2024cold}. Since the subsequent reaction with acetylene commences immediately after the formation of \textbf{P-1}, complete thermalization of the intermediate prior to reaction cannot be assumed. Instead, the reacting ion population is expected to possess a distribution of internal energies that evolves through collisions with the helium buffer gas during the course of the experiment. Consequently, \textcolor{black}{the residual internal excitation of \textbf{P-1} contribute to the observed reaction kinetics}. Indeed, recent experimental studies have shown that internal excitation can either enhance or suppress ion–molecule reaction rate coefficients, depending on the underlying reaction dynamics~\citep{jimenez2025ion, hahn2024opposite, markus2020vibrational}.

\par The reported reaction mechanism (Fig.~\ref{fig4}) is likely to play a significant role in interstellar environments, since the formation of the bicyclic end product proceeds via an exothermic pathway ($\Delta H =-2.73$ eV).
The measured radiative association \textcolor{black}{rate coefficient} ((1.3$\pm$0.3) $\times$ 10$^{-10}$ cm$^3$ s$^{-1}$) for the m/z 106 species, comparable with other astrochemically relevant ion-molecule reactions (Table \ref{tab:2}), shows its potential for formation and could lead to the growth of large nitrogen-bearing PAHs under appropriate astronomical conditions.
Even though the experiments were conducted at 295 K, the comparable reaction \textcolor{black}{rate coefficients} (Table \ref{tab:2}) indicate that these reactions can proceed efficiently even under cold conditions. This suggests they could play a significant role in the chemistry of the cold ISM.

{\color{black}
\par Figure \ref{fig5} provides a schematic overview of the reaction channels studied in this work.
Of particular interest is the final reaction product, \textbf{P-2}, in which the nitrogen atom is incorporated into the aromatic carbon framework to form an endocyclic N-PAHs. Such endocyclic N-PAHs have recently been proposed as promising carriers of the characteristic 6.2 $\mu$m aromatic infrared band, owing to their ability to reproduce the \ce{C=C} stretching vibration observed in the infrared emission spectra of the interstellar medium~\citep{hudgins2005variations,canelo2018variations}. 
To examine this possibility, Fig.~\ref{IR_comparision} compares the calculated infrared absorption spectrum of \textbf{P-2} with a typical interstellar infrared emission spectrum, together with the calculated absorption spectra of common PAHs naphthalene$^+$ and coronene$^+$. The calculated frequencies were scaled with a factor 0.9670 for all the vibrational modes, and each vibrational transition was broadened using a Lorentzian profile with a full width at half maximum of 16 cm$^{-1}$~\citep{boersma2014nasa, hua2025gas}. The astronomical spectrum was adapted from~\citet{sloan2003uniform}, while the infrared spectra of naphthalene$^+$ and coronene$^+$ were obtained from the NASA Ames PAH IR Spectroscopic Database~\citep{ricca2026nasa, mattioda2020nasa, boersma2014nasa,bauschlicher2018nasa}.
\par As shown in Fig.~\ref{IR_comparision}, the \ce{C=C} stretching band of coronene$^+$ (6.4 $\mu$m) is shifted toward shorter wavelengths compared to that of naphthalene$^+$ (6.6 $\mu$m), consistent with the well-established trend that the position of this vibrational mode depends on PAHs size~\citep{hudgins1999spacing}.
In contrast, the incorporation of a nitrogen atom into the small aromatic ring produces a substantially larger shift, placing the corresponding \ce{C=C} stretching band of \textbf{P-2} at 6.16 $\mu$m compared to the fully carbon species-analog naphthalene (6.6 $\mu$m), and in close agreement with the observed 6.2 $\mu$m emission band. This result highlights the significant influence of endocyclic nitrogen incorporation on the vibrational properties of aromatic molecules and supports the inclusion of endocyclic N-PAHs in astrochemical models aimed at explaining the origin of the 6.2 $\mu$m interstellar emission feature.
Future laboratory infrared spectroscopic measurements of these species will provide an important benchmark for validating these theoretical predictions and assessing their potential role in the chemistry of the interstellar medium.
}

\par The reaction between pyrimidine and acetylene is especially intriguing as mass spectroscopy of Titan's atmosphere with Cassini's INMS has identified a peak at mass 81, which could correspond to protonated pyrimidine (\ce{C4H5N2^{+}})~\citep{nixon2024composition,vuitton2007ion}.
Recently, Nixon et al. ~\citep{nixon2020detection} established a 2-$\sigma$ upper limit of 0.85 ppb for pyrimidine in Titan's atmosphere using Atacama Large Millimeter/submillimeter Array (ALMA) data, under the assumption of uniform abundance profile above 300 km altitude.
Even without direct detection, theoretical work suggests multiple pathways for pyrimidine formation in Titan's atmosphere~\citep{jeilani2015acetylene}.
Laboratory investigation further demonstrates that acetylene cation can react with hydrogen cyanide (\ce{HCN}) and form pyrimidine cations~\citep{hamid2014evidence}.
Both Acetylene and \ce{HCN} are present in Titan's atmosphere at a mole fraction of $2.8\times10^{-4}$ and $2.0\times10^{-4}$ respectively~\citep{waite2005ion,vuitton2007ion}.
More specifically, in its ionosphere at an altitude of approximately 1100 km, where the temperatures range from 100 K to 200 K and neutral densities are of the order $10^{10}$ cm$^{-3}$,
can initiate the ion-molecule reaction that synthesizes pyrimidine, which then could subsequently react with acetylene to form larger molecules.
Larger species up to m/z 400, have been detected in Titan's atmosphere~\citep{ali2015organic}, including highly abundant ions in the m/z 100 -- 150 range~\citep{crary2009heavy,haythornthwaite2021heavy}.
Our study shows the efficient formation of an intermediate ion with m/z 106 and a final product at m/z 131 with a \textcolor{black}{rate coefficient} of $(1.3\pm0.3)\times 10^{-10}$ cm$^3$ s$^{-1}$ and $(2.2\pm0.9)\times 10^{-12}$ cm$^3$ s$^{-1}$ respectively.
These large molecules eventually solidify into organic haze particles.
As these particles drift downwards and grow, they are believed to form the foundation of stratospheric aerosols, creating a global haze layer in Titan's atmosphere~\citep{lavvas2013aerosol,waite2007process,sagan1979tholins}.
This haze is responsible for giving Titan its characteristic golden appearance at visible wavelengths.

\par These mechanisms are not limited to Titan.
Recent thermal emission measurement of Pluto and Charon, obtained separately using the James Webb Space Telescope (JWST) Mid-Infrared Instrument (MIRI), reveal that Pluto’s atmospheric haze includes organic particles similar to those on Titan, along with hydrocarbon and nitrile ices ~\citep{bertrand2025evidence, lellouch2025pluto}.
These compositional similarities suggest that similar pathways eventually lead to the formation of N-PAHs in Pluto's atmosphere.
Analogous haze formation mechanisms are also likely occurring within Triton's atmospheric layers ~\citep{broadfoot1989ultraviolet}, indicating that such processes are a widespread phenomenon among nitrogen-rich atmospheres~\citep{yang2024low,gladstone2019new}.

\par Furthermore, the study of pyrimidine reactions is not solely a pursuit of interstellar chemistry; it also provides direct insights into astrobiology.
Pyrimidine's fundamental role in nucleic acids, its presence in meteorites~\citep{oba2022identifying,herd2011origin}, and direct experimental evidence of its gas-phase formation from simpler precursors~\citep{hamid2014evidence} suggest plausible abiotic pathways for synthesizing life's fundamental building blocks.
This research, therefore, contributes to understanding the exogenous delivery of prebiotic material to early Earth~\citep{anslow2023can}, a critical hypothesis in the origin of life.

\par In summary, continued investigation of ion-molecule reactions involving N-PAHs is essential to unveil novel reaction pathways and identify unknown molecular species.
Although no nitrogen-bearing bicyclic aromatic molecules have yet been detected in the interstellar medium or (exo)planetary atmospheres, our findings suggest that these species are promising candidates for future astronomical searches, particularly in Titan’s nitrogen-rich atmosphere.

\section{Conclusion}
Our experimental investigation demonstrates that pyrimidine ions react efficiently and sequentially with acetylene in the gas phase to form covalently bonded, nitrogen-bearing PAH cations at 295 K.
The measured reaction \textcolor{black}{rate coefficients} and their exothermic nature indicate that these pathways are not only plausible but also assured to significantly contribute to the molecular diversity observed in astronomical environments, such as Titan's nitrogen-rich atmosphere and the ISM.
Furthermore, experimentally observed reaction mechanisms augmented by electronic structure calculations provide crucial insights for astrochemical models, allowing for more accurate simulations of ion-molecule chemistry across a range of temperatures and pressures.
However, further investigations, particularly at lower temperatures and spectroscopic probing of intermediate and final products, will undoubtedly deepen our understanding of these processes and their far-reaching implications for astrochemistry and the origin of the molecular complexity in the universe.

\vspace{0.5cm}

\begin{acknowledgments}
    This research work was supported by the Department of Science and Technology, India, through Grant No. CRG/2022/003516 (G.A.). K.R.N. thanks IIT Madras for the financial and infrastructural support through the start-up grant (NFSC).
\end{acknowledgments}

\begin{contribution}
    S.S.P. and G.A. conceived the experiments. S.S.P. planned and performed all the experiments. P.T. and Y.L. supported S.S.P. during the data collection. S.G., R.A.M., and K.R.N. planned and performed the electronic structure calculations. S.S.P. analyzed the data after discussing with all the co-authors. S.S.P. wrote the manuscript with inputs received from the co-authors.
\end{contribution}

\bibliography{pyrimidine}{}

@article{behera202422,
  title={A 22-pole radiofrequency ion trap setup for laboratory astrophysical studies},
  author={Behera, Nihar Ranjan and Dutta, Saurav and Chacko, Roby and Barik, Saroj and Aravind, G},
  journal={Review of Scientific Instruments},
  volume={95},
  number={1},
  pages = {013201},
  year={2024},
  publisher={AIP Publishing}
}

@article{nguyen1997protonation,
  title={Protonation sites in pyrimidine and pyrimidinamines in the gas phase},
  author={Nguyen, Viet Q and Turecek, Franti{\v{s}}ek},
  journal={Journal of the American Chemical Society},
  volume={119},
  number={9},
  pages={2280--2290},
  year={1997},
  publisher={ACS Publications}
}

@article{rap2022low,
  title={Low-temperature nitrogen-bearing polycyclic aromatic hydrocarbon formation routes validated by infrared spectroscopy},
  author={Rap, Dani{\"e}l B and Schrauwen, Johanna GM and Marimuthu, Aravindh N and Redlich, Britta and Br{\"u}nken, Sandra},
  journal={Nature Astronomy},
  volume={6},
  number={9},
  pages={1059--1067},
  year={2022},
  publisher={Nature Publishing Group UK London}
}

@article{payra2025dissociative,
  title={Dissociative ionization of pyrazine: A pathway to reactive interstellar ions},
  author={Payra, Siddhartha S and Lenka, Yash and Thakkar, Pratikkumar and Behera, Nihar Ranjan and Dutta, Saurav and Aravind, G},
  journal={The Journal of Chemical Physics},
  volume={163},
  number={23},
  pages={234303},
  year={2025},
  publisher={AIP Publishing}
}

@article{payra2026wavelength,
  title={Wavelength-dependent photofragmentation of pyrazine},
  author={Payra, Siddhartha S and Thakkar, Pratikkumar and Lenka, Yash and Aravind, G},
  journal={Scientific Reports},
  volume={16},
  number={1},
  pages={12113},
  year={2026},
  publisher={Nature Publishing Group UK London}
}

@article{soliman2015growth,
  title={Growth kinetics and formation mechanisms of complex organics by sequential reactions of acetylene with ionized aromatics},
  author={Soliman, Abdel-Rahman and Attah, Isaac K and Hamid, Ahmed M and El-Shall, M Samy},
  journal={International Journal of Mass Spectrometry},
  volume={377},
  pages={139--151},
  year={2015},
  publisher={Elsevier}
}

@article{rap2024noncovalent,
  title={Noncovalent interactions steer the formation of polycyclic aromatic hydrocarbons},
  author={Rap, Dani{\"e}l B and Schrauwen, Johanna GM and Redlich, Britta and Brunken, Sandra},
  journal={Journal of the American Chemical Society},
  volume={146},
  number={33},
  pages={23022--23033},
  year={2024},
  publisher={ACS Publications}
}

@article{shiels2021reactivity,
  title={Reactivity trends in the gas-phase addition of acetylene to the N-protonated aryl radical cations of pyridine, aniline, and benzonitrile},
  author={Shiels, Oisin J and Kelly, PD and Bright, Cameron C and Poad, Berwyck LJ and Blanksby, Stephen J and Da Silva, Gabriel and Trevitt, Adam J},
  journal={Journal of the American Society for Mass Spectrometry},
  volume={32},
  number={2},
  pages={537--547},
  year={2021},
  publisher={ACS Publications}
}

@article{soliman2013formation,
  title={Formation of nitrogen-containing polycyclic cations by gas-phase and intracluster reactions of acetylene with the pyridinium and pyrimidinium ions},
  author={Soliman, Abdel-Rahman and Hamid, Ahmed M and Attah, Isaac and Momoh, Paul and El-Shall, M Samy},
  journal={Journal of the American Chemical Society},
  volume={135},
  number={1},
  pages={155--166},
  year={2013},
  publisher={ACS Publications}
}

@article{momoh2008formation,
  title={Formation of complex organics from acetylene catalyzed by ionized benzene},
  author={Momoh, Paul O and Soliman, Abdel-Rahman and Meot-Ner, Michael and Ricca, Alessandra and El-Shall, M Samy},
  journal={Journal of the American Chemical Society},
  volume={130},
  number={39},
  pages={12848--12849},
  year={2008},
  publisher={ACS Publications}
}

@article{ozturk1989reactions,
  title={Reactions of C5H3+ and C5H5+ ions with acetylene and diacetylene},
  author={Ozturk, Feza and Moini, Mehdi and Brill, Fred W and Eyler, John R and Buckley, Thomas J and Lias, Sharon G and Ausloos, Pierre J},
  journal={The Journal of Physical Chemistry},
  volume={93},
  number={10},
  pages={4038--4044},
  year={1989},
  publisher={ACS Publications}
}

@article{soliman2012formation,
  title={Formation of complex organics in the gas phase by sequential reactions of acetylene with the phenylium ion},
  author={Soliman, Abdel-Rahman and Hamid, Ahmed M and Momoh, Paul O and El-Shall, M Samy and Taylor, Danielle and Gallagher, Lauren and Abrash, Samuel A},
  journal={The Journal of Physical Chemistry A},
  volume={116},
  number={36},
  pages={8925--8933},
  year={2012},
  publisher={ACS Publications}
}

@article{nixon2020detection,
  title={Detection of Cyclopropenylidene on Titan with ALMA},
  author={Nixon, Conor A and Thelen, Alexander E and Cordiner, Martin A and Kisiel, Zbigniew and Charnley, Steven B and Molter, Edward M and Serigano, Joseph and Irwin, Patrick GJ and Teanby, Nicholas A and Kuan, Yi-Jehng},
  journal={The Astronomical Journal},
  volume={160},
  number={5},
  pages={205},
  year={2020},
  publisher={IOP Publishing}
}

@article{vuitton2007ion,
  title={Ion chemistry and N-containing molecules in Titan's upper atmosphere},
  author={Vuitton, V and Yelle, RV and McEwan, MJ},
  journal={Icarus},
  volume={191},
  number={2},
  pages={722--742},
  year={2007},
  publisher={Elsevier}
}

@article{hamid2014evidence,
  title={Evidence for the formation of pyrimidine cations from the sequential reactions of hydrogen cyanide with the acetylene radical cation},
  author={Hamid, Ahmed M and Bera, Partha P and Lee, Timothy J and Aziz, Saadullah G and Alyoubi, Abdulrahman O and El-Shall, M Samy},
  journal={The journal of physical chemistry letters},
  volume={5},
  number={19},
  pages={3392--3398},
  year={2014},
  publisher={ACS Publications}
}

@article{jeilani2015acetylene,
  title={Acetylene as an essential building block for prebiotic formation of pyrimidine bases on Titan},
  author={Jeilani, Yassin A and Fearce, Chelesa and Nguyen, Minh Tho},
  journal={Physical Chemistry Chemical Physics},
  volume={17},
  number={37},
  pages={24294--24303},
  year={2015},
  publisher={Royal Society of Chemistry}
}

@article{waite2005ion,
  title={Ion neutral mass spectrometer results from the first flyby of Titan},
  author={Waite Jr, J Hunter and Niemann, Hasso and Yelle, Roger V and Kasprzak, Wayne T and Cravens, Thomas E and Luhmann, Janet G and McNutt, Ralph L and Ip, Wing-Huen and Gell, David and De La Haye, Virginie and others},
  journal={Science},
  volume={308},
  number={5724},
  pages={982--986},
  year={2005},
  publisher={American Association for the Advancement of Science}
}

@article{waite2007process,
  title={The process of tholin formation in Titan's upper atmosphere},
  author={Waite Jr, JH and Young, DT and Cravens, TE and Coates, AJ and Crary, FJ and Magee, B and Westlake, J},
  journal={Science},
  volume={316},
  number={5826},
  pages={870--875},
  year={2007},
  publisher={American Association for the Advancement of Science}
}

@article{sagan1979tholins,
  title={Tholins: Organic chemistry of interstellar grains and gas},
  author={Sagan, Carl and Khare, BN},
  journal={Nature},
  volume={277},
  number={5692},
  pages={102--107},
  year={1979},
  publisher={Nature Publishing Group UK London}
}

@article{crary2009heavy,
  title={Heavy ions, temperatures and winds in Titan's ionosphere: Combined Cassini CAPS and INMS observations},
  author={Crary, FJ and Magee, BA and Mandt, K and Waite Jr, JH and Westlake, J and Young, DT},
  journal={Planetary and Space Science},
  volume={57},
  number={14-15},
  pages={1847--1856},
  year={2009},
  publisher={Elsevier}
}

@article{haythornthwaite2021heavy,
  title={Heavy positive ion groups in titan’s ionosphere from cassini plasma spectrometer IBS observations},
  author={Haythornthwaite, Richard P and Coates, Andrew J and Jones, Geraint H and Wellbrock, Anne and Waite, J Hunter and Vuitton, V{\'e}ronique and Lavvas, Panayotis},
  journal={The Planetary Science Journal},
  volume={2},
  number={1},
  pages={26},
  year={2021},
  publisher={IOP Publishing}
}

@article{ali2015organic,
  title={Organic chemistry in Titan׳s upper atmosphere and its astrobiological consequences: I. Views towards Cassini plasma spectrometer (CAPS) and ion neutral mass spectrometer (INMS) experiments in space},
  author={Ali, AECS and Sittler Jr, EC and Chornay, D and Rowe, BR and Puzzarini, CRISTINA},
  journal={Planetary and Space Science},
  volume={109},
  pages={46--63},
  year={2015},
  publisher={Elsevier}
}

@article{hudgins2005variations,
  title={Variations in the peak position of the 6.2 $\mu$m interstellar emission feature: A tracer of N in the interstellar polycyclic aromatic hydrocarbon population},
  author={Hudgins, Douglas M and Bauschlicher Jr, Charles W and Allamandola, LJ},
  journal={The Astrophysical Journal},
  volume={632},
  number={1},
  pages={316},
  year={2005},
  publisher={IOP Publishing}
}

@article{canelo2018variations,
  title={Variations in the 6.2 $\mu$m emission profile in starburst-dominated galaxies: a signature of polycyclic aromatic nitrogen heterocycles (PANHs)?},
  author={Canelo, Carla M and Fria{\c{c}}a, Am{\^a}ncio CS and Sales, Dinalva A and Pastoriza, Miriani G and Ruschel-Dutra, Daniel},
  journal={Monthly Notices of the Royal Astronomical Society},
  volume={475},
  number={3},
  pages={3746--3763},
  year={2018},
  publisher={Oxford University Press}
}

@article{kumar2019effects,
  title={Effects of residual water in a linear quadrupole ion trap on the protonation sites of 4-aminobenzoic acid},
  author={Kumar, Rashmi and Yerabolu, Ravikiran and Kentt\"amaa, Hilkka I},
  journal={Journal of the American Society for Mass Spectrometry},
  volume={31},
  number={1},
  pages={124--131},
  year={2019},
  publisher={ACS Publications}
}

@article{li2020spitzer,
  title={Spitzer’s perspective of polycyclic aromatic hydrocarbons in galaxies},
  author={Li, Aigen},
  journal={Nature Astronomy},
  volume={4},
  number={4},
  pages={339--351},
  year={2020},
  publisher={Nature Publishing Group UK London}
}

@article{allamandola2021pah,
  title={PAH Spectroscopy from 1 to 5 $\mu$m},
  author={Allamandola, LJ and Boersma, C and Lee, TJ and Bregman, JD and Temi, P},
  journal={The Astrophysical Journal Letters},
  volume={917},
  number={2},
  pages={L35},
  year={2021},
  publisher={IOP Publishing}
}

@article{mattioda2008near,
  title={Near-infrared spectroscopy of nitrogenated polycyclic aromatic hydrocarbon cations from 0.7 to 2.5 $\mu$m},
  author={Mattioda, Andrew L and Rutter, Lindsay and Parkhill, John and Head-Gordon, Martin and Lee, Timothy J and Allamandola, Louis J},
  journal={The Astrophysical Journal},
  volume={680},
  number={2},
  pages={1243},
  year={2008},
  publisher={IOP Publishing}
}

@article{peeters2002rich,
  title={The rich 6 to 9 m spectrum of interstellar PAHs},
  author={Peeters, E and Hony, S and Van Kerckhoven, C and Tielens, AGGM and Allamandola, LJ and Hudgins, DM and Bauschlicher, CW},
  journal={Astronomy \& Astrophysics},
  volume={390},
  number={3},
  pages={1089--1113},
  year={2002},
  publisher={EDP Sciences}
}

@article{wenzel2025discovery,
  title={Discovery of the Seven-ring Polycyclic Aromatic Hydrocarbon Cyanocoronene (C24H11CN) in GOTHAM Observations of TMC-1},
  author={Wenzel, Gabi and Gong, Siyuan and Xue, Ci and Changala, P Bryan and Holdren, Martin S and Speak, Thomas H and Stewart, D Archie and Fried, Zachary TP and Willis, Reace HJ and Bergin, Edwin A and others},
  journal={The Astrophysical Journal Letters},
  volume={984},
  number={1},
  pages={L36},
  year={2025},
  publisher={IOP Publishing}
}

@article{wenzel2025detections,
  title={Detections of interstellar aromatic nitriles 2-cyanopyrene and 4-cyanopyrene in TMC-1},
  author={Wenzel, Gabi and Speak, Thomas H and Changala, P Bryan and Willis, Reace HJ and Burkhardt, Andrew M and Zhang, Shuo and Bergin, Edwin A and Byrne, Alex N and Charnley, Steven B and Fried, Zachary TP and others},
  journal={Nature Astronomy},
  volume={9},
  number={2},
  pages={262--270},
  year={2025},
  publisher={Nature Publishing Group UK London}
}

@article{wenzel2024detection,
  title={Detection of interstellar 1-cyanopyrene: A four-ring polycyclic aromatic hydrocarbon},
  author={Wenzel, Gabi and Cooke, Ilsa R and Changala, P Bryan and Bergin, Edwin A and Zhang, Shuo and Burkhardt, Andrew M and Byrne, Alex N and Charnley, Steven B and Cordiner, Martin A and Duffy, Miya and others},
  journal={Science},
  volume={386},
  number={6723},
  pages={810--813},
  year={2024},
  publisher={American Association for the Advancement of Science}
}

@article{mcguire2021detection,
  title={Detection of two interstellar polycyclic aromatic hydrocarbons via spectral matched filtering},
  author={McGuire, Brett A and Loomis, Ryan A and Burkhardt, Andrew M and Lee, Kin Long Kelvin and Shingledecker, Christopher N and Charnley, Steven B and Cooke, Ilsa R and Cordiner, Martin A and Herbst, Eric and Kalenskii, Sergei and others},
  journal={Science},
  volume={371},
  number={6535},
  pages={1265--1269},
  year={2021},
  publisher={American Association for the Advancement of Science}
}

@article{anslow2023can,
  title={Can comets deliver prebiotic molecules to rocky exoplanets?},
  author={Anslow, Richard J and Bonsor, Amy and Rimmer, Paul B},
  journal={Proceedings of the Royal Society A},
  volume={479},
  number={2279},
  pages={20230434},
  year={2023},
  publisher={The Royal Society}
}

@article{oba2022identifying,
  title={Identifying the wide diversity of extraterrestrial purine and pyrimidine nucleobases in carbonaceous meteorites},
  author={Oba, Yasuhiro and Takano, Yoshinori and Furukawa, Yoshihiro and Koga, Toshiki and Glavin, Daniel P and Dworkin, Jason P and Naraoka, Hiroshi},
  journal={Nature communications},
  volume={13},
  number={1},
  pages={2008},
  year={2022},
  publisher={Nature Publishing Group UK London}
}

@article{herd2011origin,
  title={Origin and evolution of prebiotic organic matter as inferred from the Tagish Lake meteorite},
  author={Herd, Christopher DK and Blinova, Alexandra and Simkus, Danielle N and Huang, Yongsong and Tarozo, Rafael and Alexander, Conel M O’D and Gyngard, Frank and Nittler, Larry R and Cody, George D and Fogel, Marilyn L and others},
  journal={Science},
  volume={332},
  number={6035},
  pages={1304--1307},
  year={2011},
  publisher={American Association for the Advancement of Science}
}

@article{kaiser2021aromatic,
  title={An aromatic universe--A physical chemistry perspective},
  author={Kaiser, Ralf I and Hansen, Nils},
  journal={The Journal of Physical Chemistry A},
  volume={125},
  number={18},
  pages={3826--3840},
  year={2021},
  publisher={ACS Publications}
}

@article{sandford2020prebiotic,
  title={Prebiotic astrochemistry and the formation of molecules of astrobiological interest in interstellar clouds and protostellar disks},
  author={Sandford, Scott A and Nuevo, Michel and Bera, Partha P and Lee, Timothy J},
  journal={Chemical reviews},
  volume={120},
  number={11},
  pages={4616--4659},
  year={2020},
  publisher={ACS Publications}
}

@article{doddipatla2021low,
  title={Low-temperature gas-phase formation of indene in the interstellar medium},
  author={Doddipatla, Srinivas and Galimova, Galiya R and Wei, Hongji and Thomas, Aaron M and He, Chao and Yang, Zhenghai and Morozov, Alexander N and Shingledecker, Christopher N and Mebel, Alexander M and Kaiser, Ralf I},
  journal={Science advances},
  volume={7},
  number={1},
  pages={eabd4044},
  year={2021},
  publisher={American Association for the Advancement of Science}
}

@article{prasad1983uv,
  title={UV radiation field inside dense clouds-Its possible existence and chemical implications},
  author={Prasad, Sheo S and Tarafdar, Shankar P},
  journal={Astrophysical Journal, Part 1 (ISSN 0004-637X), vol. 267, April 15, 1983, p. 603-609. NASA-supported research.},
  volume={267},
  pages={603--609},
  year={1983}
}

@article{lavvas2013aerosol,
  title={Aerosol growth in Titan’s ionosphere},
  author={Lavvas, Panayotis and Yelle, Roger V and Koskinen, Tommi and Bazin, Axel and Vuitton, V{\'e}ronique and Vigren, Erik and Galand, Marina and Wellbrock, Anne and Coates, Andrew J and Wahlund, Jan-Erik and others},
  journal={Proceedings of the National Academy of Sciences},
  volume={110},
  number={8},
  pages={2729--2734},
  year={2013},
  publisher={National Academy of Sciences}
}

@article{nixon2024composition,
  title={The composition and chemistry of Titan’s atmosphere},
  author={Nixon, Conor A},
  journal={ACS Earth and Space Chemistry},
  volume={8},
  number={3},
  pages={406--456},
  year={2024},
  publisher={ACS Publications}
}

@article{broadfoot1989ultraviolet,
  title={Ultraviolet spectrometer observations of Neptune and Triton},
  author={Broadfoot, AL and Atreya, SK and Bertaux, JL and Blamont, JE and Dessler, AJ and Donahue, TM and Forrester, WT and Hall, DT and Herbert, F and Holberg, JB and others},
  journal={Science},
  volume={246},
  number={4936},
  pages={1459--1466},
  year={1989},
  publisher={American Association for the Advancement of Science}
}

@article{lellouch2025pluto,
  title={Pluto’s atmosphere gas and haze composition from JWST/MIRI spectroscopy},
  author={Lellouch, Emmanuel and Wong, I and Lavvas, Panayotis and Bertrand, Tanguy and Villanueva, G and Stansberry, J and Holler, B and Pinilla-Alonso, N and Merlin, Fr{\'e}d{\'e}ric and Souza-Feliciano, AC and others},
  journal={Astronomy \& Astrophysics},
  volume={696},
  pages={A147},
  year={2025},
  publisher={EDP Sciences}
}

@article{yang2024low,
  title={Low-temperature formation of pyridine and (iso) quinoline via neutral--neutral reactions},
  author={Yang, Zhenghai and He, Chao and Goettl, Shane J and Mebel, Alexander M and Velloso, Paulo FG and Alves, M{\'a}rcio O and Galv{\~a}o, Breno RL and Loison, Jean-Christophe and Hickson, Kevin M and Dobrijevic, Michel and others},
  journal={Nature Astronomy},
  volume={8},
  number={7},
  pages={856--864},
  year={2024},
  publisher={Nature Publishing Group UK London}
}

@article{gladstone2019new,
  title={New Horizons observations of the atmosphere of Pluto},
  author={Gladstone, G Randall and Young, Leslie A},
  journal={Annual Review of Earth and Planetary Sciences},
  volume={47},
  number={1},
  pages={119--140},
  year={2019},
  publisher={Annual Reviews}
}

@article{bertrand2025evidence,
  title={Evidence of haze control of Pluto’s atmospheric heat balance from JWST/MIRI thermal light curves},
  author={Bertrand, Tanguy and Lellouch, Emmanuel and Holler, Bryan and Stansberry, John and Wong, Ian and Zhang, Xi and Lavvas, Panayotis and Dufaux, Elodie and Merlin, Frederic and Villanueva, Geronimo and others},
  journal={Nature Astronomy},
  volume={9},
  number={9},
  pages={1300--1308},
  year={2025},
  publisher={Nature Publishing Group UK London}
}

@misc{g16,
author={M. J. 
and G. W. Trucks and H. B. Schlegel and G. E. Scuseria and M. A. Robb and J. R. Cheeseman and G. Scalmani and V. Barone and G. A. Petersson and H. Nakatsuji and X. Li and M. Caricato and A. V. Marenich and J. Bloino and B. G. Janesko and R. Gomperts and B. Mennucci and H. P. Hratchian and J. V. Ortiz and A. F. Izmaylov and J. L. Sonnenberg and D. Williams-Young and F. Ding and F. Lipparini and F. Egidi and J. Goings and B. Peng and A. Petrone and T. Henderson and D. Ranasinghe and V. G. Zakrzewski and J. Gao and N. Rega and G. Zheng and W. Liang and M. Hada and M. Ehara and K. Toyota and R. Fukuda and J. Hasegawa and M. Ishida and T. Nakajima and Y. Honda and O. Kitao and H. Nakai and T. Vreven and K. Throssell and Montgomery, {Jr.}, J. A. and J. E. Peralta and F. Ogliaro and M. J. Bearpark and J. J. Heyd and E. N. Brothers and K. N. Kudin and V. N. Staroverov and T. A. Keith and R. Kobayashi and J. Normand and K. Raghavachari and A. P. Rendell and J. C. Burant and S. S. Iyengar and J. Tomasi and M. Cossi and J. M. Millam and M. Klene and C. Adamo and R. Cammi and J. W. Ochterski and R. L. Martin and K. Morokuma and O. Farkas and J. B. Foresman and D. J. Fox},
title={Gaussian˜16 {R}evision {C}.01},
year={2016},
note={Gaussian Inc. Wallingford CT}
}

@article{sloan2003uniform,
  title={A uniform database of 2.4-45.4 micron spectra from the Infrared Space Observatory Short Wavelength Spectrometer},
  author={Sloan, GC and Kraemer, Kathleen E and Price, Stephan D and Shipman, Russell F},
  journal={The Astrophysical Journal Supplement Series},
  volume={147},
  number={2},
  pages={379--401},
  year={2003}
}

@article{ricca2026nasa,
  title={The NASA Ames PAH IR Spectroscopic Database: Computational Version 4.00, Software Tools, Website, and Documentation},
  author={Ricca, A and Boersma, C and Maragkoudakis, A and Roser, JE and Shannon, Matthew J and Allamandola, LJ and Bauschlicher Jr, Charles W},
  journal={The Astrophysical Journal Supplement Series},
  volume={282},
  number={1},
  pages={7},
  year={2026},
  publisher={The American Astronomical Society}
}

@article{bauschlicher2018nasa,
  title={The NASA Ames PAH IR spectroscopic database: Computational version 3.00 with updated content and the introduction of multiple scaling factors},
  author={Bauschlicher Jr, Charles W and Ricca, A and Boersma, C and Allamandola, LJ},
  journal={The Astrophysical Journal Supplement Series},
  volume={234},
  number={2},
  pages={32},
  year={2018},
  publisher={The American Astronomical Society}
}

@article{boersma2014nasa,
  title={The NASA Ames PAH IR spectroscopic database version 2.00: updated content, web site, and on (off) line tools},
  author={Boersma, C and Bauschlicher Jr, CW and Ricca, A and Mattioda, AL and Cami, J and Peeters, E and de Armas, F S{\'a}nchez and Saborido, G Puerta and Hudgins, DM and Allamandola, LJ},
  journal={The Astrophysical Journal Supplement Series},
  volume={211},
  number={1},
  pages={8},
  year={2014},
  publisher={The American Astronomical Society}
}

@article{mattioda2020nasa,
  title={The NASA Ames PAH IR spectroscopic database: The laboratory spectra},
  author={Mattioda, AL and Hudgins, DM and Boersma, C and Bauschlicher Jr, CW and Ricca, A and Cami, J and Peeters, E and de Armas, F S{\'a}nchez and Saborido, G Puerta and Allamandola, LJ},
  journal={The Astrophysical Journal Supplement Series},
  volume={251},
  number={2},
  pages={22},
  year={2020},
  publisher={The American Astronomical Society}
}

@article{hua2025gas,
  title={Gas-phase formation and photochemistry of large cyano-containing polycyclic aromatic hydrocarbon clusters},
  author={Hua, Lijun and Wu, Xianghuang and Zhang, Congcong and Zhao, Nian and Ye, Ting and Xiao, Huaping and Zhen, Junfeng and Yang, Xuejuan},
  journal={Astronomy \& Astrophysics},
  volume={700},
  pages={A31},
  year={2025},
  publisher={EDP Sciences}
}

@article{hudgins1999spacing,
  title={The spacing of the interstellar 6.2 and 7.7 micron emission features as an indicator of polycyclic aromatic hydrocarbon size},
  author={Hudgins, Douglas M and Allamandola, LJ},
  journal={The Astrophysical Journal Letters},
  volume={513},
  number={1},
  pages={L69--L73},
  year={1999}
}

@article{jimenez2025ion,
  title={Ion-kinetic-energy sampling in a 22-pole trap using ring-electrode evaporation},
  author={Jim{\'e}nez-Redondo, Miguel and Gerlich, Dieter and Caselli, Paola and Jusko, Pavol},
  journal={Physical Review A},
  volume={111},
  number={1},
  pages={013120},
  year={2025},
  publisher={APS}
}

@article{hahn2024opposite,
  title={Opposite effects of the rotational and translational energy on the rates of ion-molecule reactions near 0 K: the D 2++ NH 3 and D 2++ ND 3 reactions},
  author={Hahn, Rapha{\"e}l and Schlander, David and Zhelyazkova, Valentina and Merkt, Fr{\'e}d{\'e}ric},
  journal={Physical Review X},
  volume={14},
  number={1},
  pages={011034},
  year={2024},
  publisher={APS}
}

@article{markus2020vibrational,
  title={Vibrational excitation hindering an ion-molecule reaction: The c-C 3 H 2+- H 2 collision complex},
  author={Markus, Charles R and Asvany, Oskar and Salomon, Thomas and Schmid, Philipp C and Br{\"u}nken, Sandra and Lipparini, Filippo and Gauss, J{\"u}rgen and Schlemmer, Stephan},
  journal={Physical Review Letters},
  volume={124},
  number={23},
  pages={233401},
  year={2020},
  publisher={APS}
}

@article{jusko2024cold,
  title={Cold CAS ion trap--22 pole trap with ring electrodes for astrochemistry},
  author={Jusko, Pavol and Jim{\'e}nez-Redondo, Miguel and Caselli, Paola},
  journal={Molecular Physics},
  volume={122},
  number={1-2},
  pages={e2217744},
  year={2024},
  publisher={Taylor \& Francis}
}
\bibliographystyle{aasjournalv7}

\begin{appendix}

\section{Experimental setup}
\begin{figure*}[ht!]
    \centering
    \includegraphics[width=0.9\textwidth]{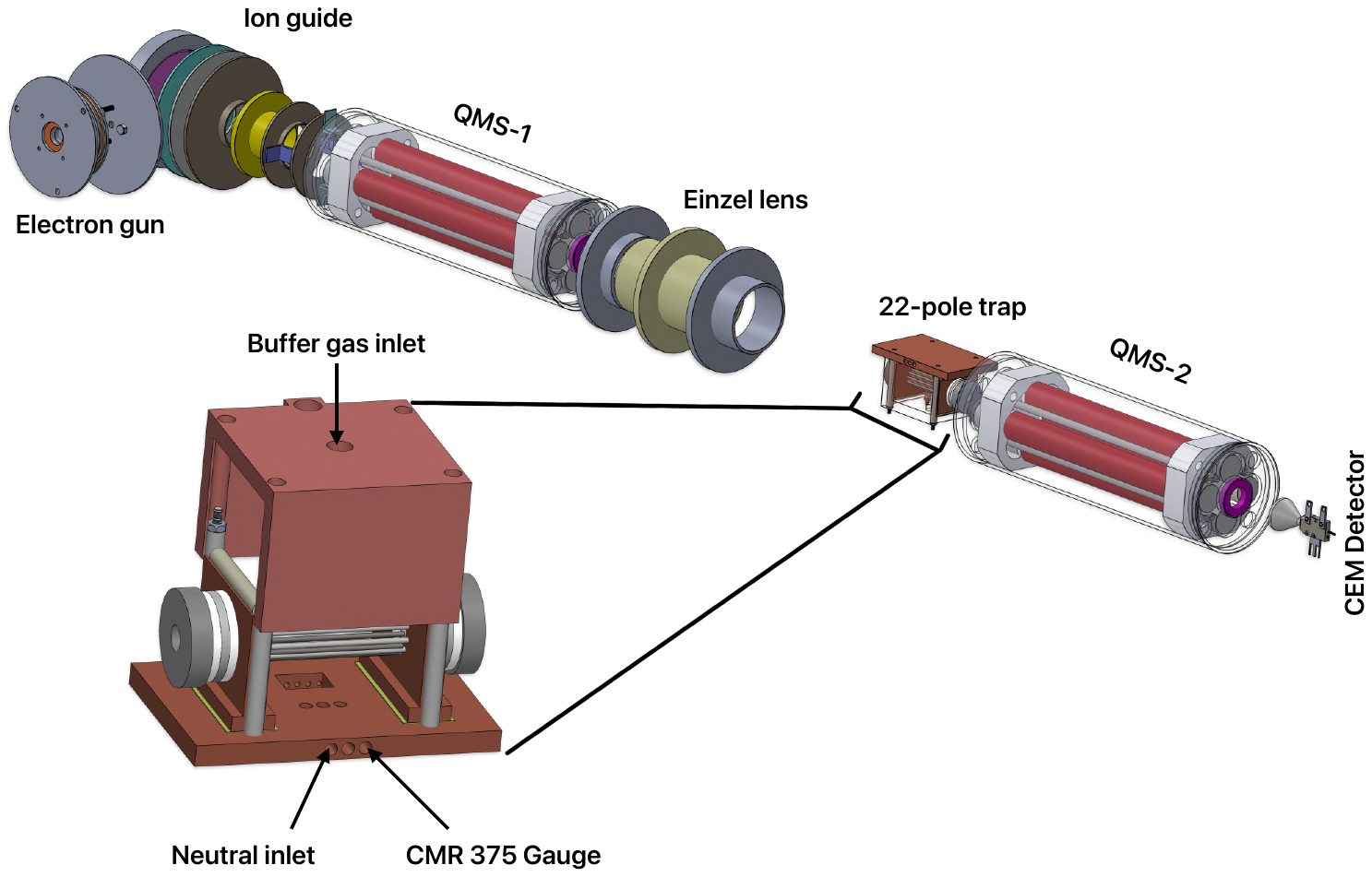}
    \caption{Schematic drawing of the modified 22-pole radio frequency ion trap setup. The ions are produced between the first two plates of the ion guide by electron impact ionization and repelled towards the QMS-1. Mass-selected ions are injected into the 22-pole trap. The trap is equipped with neutral and pulsed buffer gas input ports. A neutral density measurement gauge (Pfeiffer CMR 375) is also connected to the trap volume. After a specified trapping time, the ions are mass-analyzed by QMS-2 and counted in the CEM detector.
    \label{exp_setup}}
\end{figure*}

\section{ODE model for rate coefficient calculation}
We fitted the experimental kinetic data using the following rate equation (with the initial number of ions and the reaction rate coefficients as free parameters) to determine \textcolor{black}{rate coefficients}.
\[
\frac{dN_{\mathrm{C_4H_4N_2^+}}}{dt} = -\tilde{k}_{\mathrm{RA}} N_{\mathrm{C_4H_4N_2^+}}
\]
\[
\frac{dN_{\mathrm{C_4H_5N_2^+}}}{dt} = -\tilde{k}_1 N_{\mathrm{C_4H_5N_2^+}}
\]
\[
\frac{dN_{\mathrm{C_6H_6N_2^+}}}{dt}=+\tilde{k}_{\mathrm{RA}}N_{\mathrm{C_4H_4N_2^+}}+\tilde{k}_1 N_{\mathrm{C_4H_5N_2^+}}
\]
\[
\frac{dN_{\mathrm{C_6H_6N_2^+}}}{dt} = -\tilde{k}_2 N_{\mathrm{C_6H_6N_2^+}}
\]
Where $N_{\mathrm{x}^+}$ is the detected number of x$^+$ ion and the effective rate coefficient, $\tilde{k}_i$, is given by
\[
\tilde{k}_i=k_i[C_2H_2],
\]
where [\ce{C2H2}] is the concentration of neutral \ce{C2H2} within the trap and $k_i$ is the rate coefficient.

\section{Wide range mass spectrum of the reaction of pyrimidine$^+$ with acetylene}

\begin{figure*}[ht!]
    \centering
    \includegraphics[width=0.9\textwidth]{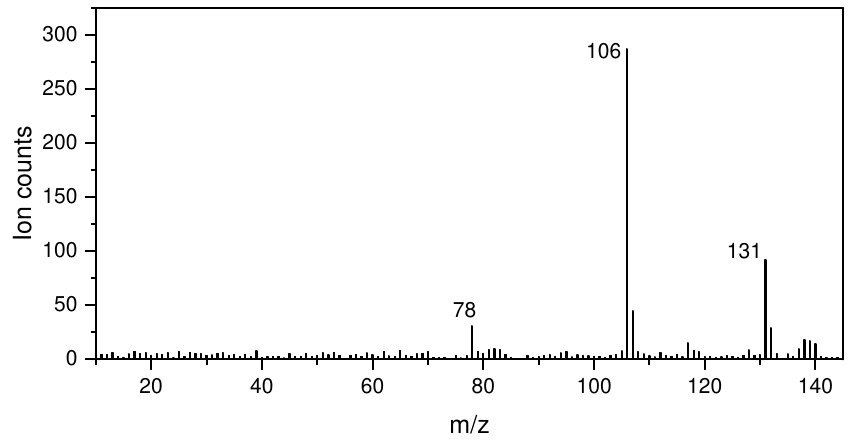}
    \caption{Wide range mass spectrum obtained for the reaction of pyrimidine$^+$ with acetylene. All the parent ions react with the acetylene to form the primary product at m/z 106, which subsequently undergoes a second acetylene addition to yield the ion at m/z 131. Very weak signals are observed at m/z 138 and 139. However, their intensities remain constant with storage time. owing to their negligible abundance and lack of temporal evolution, these peaks were excluded from the kinetic analysis.
    \label{wide_MS}}
\end{figure*}

\newpage
\FloatBarrier

\section{Transition state geometries and the imaginary frequencies}
\begin{table}[ht!]
\centering
\renewcommand{\arraystretch}{1.2}

\begin{tabular}{cccc}
\includegraphics[width=0.22\textwidth]{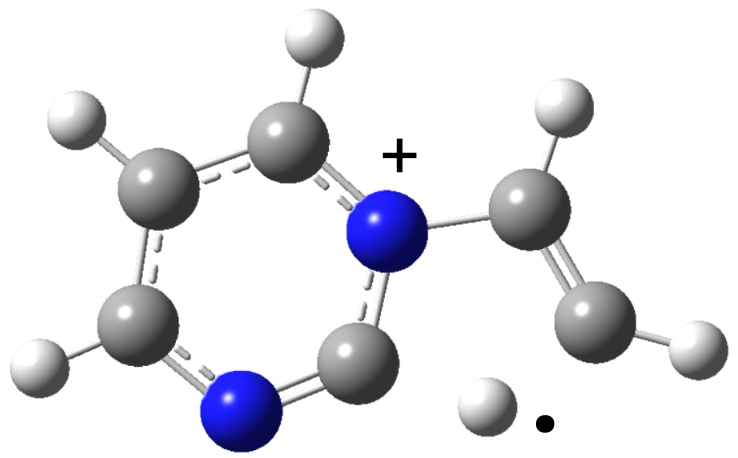} &
\includegraphics[width=0.22\textwidth]{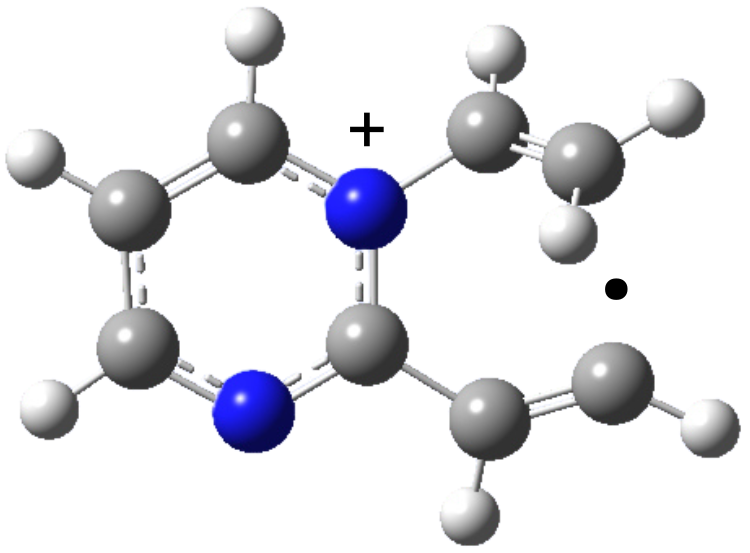} &
\includegraphics[width=0.22\textwidth]{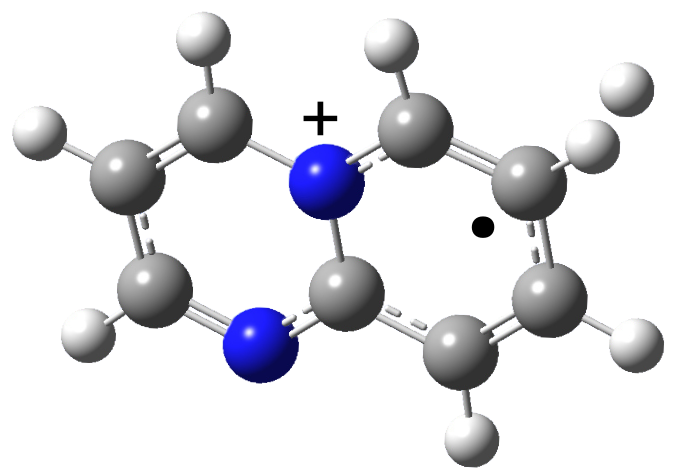} &
\\[0.3cm]

TS-1 & TS-2 & TS-3 & \\

2003.88i cm$^{-1}$ &
208.43i cm$^{-1}$ &
848.53i cm$^{-1}$ &
\\
\end{tabular}

\label{tab:TS_freq}
\end{table}

\FloatBarrier
\section{Atomic coordinates (\AA) of selected structures in the reaction profile diagram (Fig. 5 in the main text)}

\begin{table*}[ht!]
\caption{Atomic Coordinates (\AA) for INT-1, TS-1 and TS-2}
\resizebox{\textwidth}{!}{%
\begin{tabular}{ccccc|ccccc|ccccc}
\hline
\hline

\multicolumn{5}{c|}{\textbf{INT-1}} &
\multicolumn{5}{c|}{\textbf{TS-1}} &
\multicolumn{5}{c}{\textbf{TS-2}} \\
\hline
Center & Atom & X & Y & Z &
Center & Atom & X & Y & Z &
Center & Atom & X & Y & Z \\
\hline


1 & C & -0.042824 & -1.058628 & -0.000002 &
1 & C & 0.191697 & -0.928568 & 0.000001 &
1 & C & 0.059103 & 0.738190 & -0.036032 \\

2 & N & -0.514418 & 0.224181 & -0.000002 &
2 & N & 0.485421 & 0.403557 & -0.000000 &
2 & N & 0.034807 & -0.639104 & -0.132132 \\

3 & C & 0.374724 & 1.249035 & -0.000001 &
3 & C & -0.512756 & 1.303001 & -0.000001 &
3 & C & 1.204496 & -1.337035 & -0.117443 \\

4 & C & 1.726701 & 0.979802 & -0.000001 &
4 & C & -1.821061 & 0.831771 & -0.000000 &
4 & C & 2.401139 & -0.685337 & 0.033742 \\

5 & C & 2.119401 & -0.357808 & 0.000001 &
5 & C & -2.028994 & -0.543622 & 0.000001 &
5 & C & 2.349602 & 0.707683 & 0.169029 \\

6 & N & 1.227709 & -1.358941 & 0.000001 &
6 & N & -0.987246 & -1.410784 & 0.000002 &
6 & N & 1.210141 & 1.385696 & 0.119587 \\

7 & H & -0.033664 & 2.251395 & -0.000000 &
7 & H & -0.247853 & 2.353670 & -0.000002 &
7 & H & 1.113306 & -2.411311 & -0.215307 \\

8 & H & -0.789509 & -1.844766 & 0.000003 &
8 & H & 1.497946 & -1.367746 & 0.000001 &
8 & H & -2.128084 & -0.752279 & 1.483335 \\

9 & H & 2.445572 & 1.788864 & -0.000001 &
9 & H & -2.649820 & 1.527654 & -0.000001 &
9 & H & 3.333306 & -1.234003 & 0.053636 \\

10 & H & 3.165793 & -0.644378 & 0.000002 &
10 & H & -3.022708 & -0.976957 & 0.000001 &
10 & H & 3.251810 & 1.294448 & 0.311803 \\

11 & C & -2.856671 & -0.423569 & 0.000002 &
11 & C & 2.585376 & -0.512315 & -0.000000 &
11 & C & -2.219301 & -1.243291 & 0.522674 \\

12 & H & -3.929871 & -0.522704 & -0.000017 &
12 & H & 3.627054 & -0.795826 & -0.000000 &
12 & H & -3.149556 & -1.759825 & 0.319747 \\

13 & C & -1.936322 & 0.502399 & 0.000003 &
13 & C & 1.924601 & 0.626273 & -0.000001 &
13 & C & -1.197061 & -1.369550 & -0.326106 \\

14 & H & -2.161417 & 1.567515 & 0.000011 &
14 & H & 2.274979 & 1.650555 & -0.000002 &
14 & H & -1.225441 & -1.947358 & -1.241763 \\

& & & & &
& & & & &
15 & C & -2.400538 & 1.178314 & -0.080809 \\

& & & & &
& & & & &
16 & H & -3.354460 & 1.678975 & -0.181678 \\

& & & & &
& & & & &
17 & C & -1.141170 & 1.560567 & -0.178568 \\

& & & & &
& & & & &
18 & H & -0.893139 & 2.607963 & -0.360877 \\

\hline
\end{tabular}
}
\label{tab:int1_ts1_ts2}
\end{table*}

\begin{table*}
\caption{Atomic Coordinates (\AA) for TS-3, P-1 and TS-1 (Input)}
\resizebox{\textwidth}{!}{%
\begin{tabular}{ccccc|ccccc|ccccc}
\hline
\hline

\multicolumn{5}{c|}{\textbf{TS-3}} &
\multicolumn{5}{c|}{\textbf{P-1}} &
\multicolumn{5}{c}{\textbf{TS-1}} \\
\multicolumn{5}{c|}{} &
\multicolumn{5}{c|}{} &
\multicolumn{5}{c}{(Input: opt=(calcfc,tight,ts,noeigen) freq ub3lyp/6-311g(d)}\\
\multicolumn{5}{c|}{} &
\multicolumn{5}{c|}{} &
\multicolumn{5}{c}{geom=connectivity int=superfinegrid)}\\
\hline


Center & Atom & X & Y & Z &
Center & Atom & X & Y & Z &
Center & Atom & X & Y & Z \\
\hline

1 & C & 0.082463 & -0.734130 & -0.009879 &
1 & C & -0.011802 & -1.039258 & -0.000001 &
1 & C & 0.200266 & -0.934388 & -0.000016 \\

2 & N & 0.044587 & 0.664740 & -0.038325 &
2 & N & -0.492336 & 0.227494 & 0.000000 &
2 & N & 0.476609 & 0.403254 & -0.000013 \\

3 & C & 1.241074 & 1.359482 & -0.027634 &
3 & C & 0.437128 & 1.225771 & 0.000001 &
3 & C & -0.510081 & 1.293683 & -0.000007 \\

4 & C & 2.418290 & 0.683189 & 0.017094 &
4 & C & 1.786886 & 0.924976 & 0.000000 &
4 & C & -1.821220 & 0.830257 & 0.000000 \\

5 & C & 2.373999 & -0.732534 & 0.051311 &
5 & C & 2.167030 & -0.415654 & -0.000001 &
5 & C & -2.035473 & -0.535411 & 0.000010 \\

6 & N & 1.243246 & -1.401134 & 0.036224 &
6 & N & 1.205607 & -1.358668 & -0.000001 &
6 & N & -0.990987 & -1.408830 & 0.000001 \\

7 & H & 1.161968 & 2.438196 & -0.053344 &
7 & H & 0.056781 & 2.239907 & 0.000002 &
7 & H & -0.257421 & 2.333254 & -0.000010 \\

8 & H & -3.267290 & 1.204531 & -0.206949 &
8 & H & -2.611058 & -1.485171 & -0.000002 &
8 & H & 1.539608 & -1.394537 & 0.000004 \\

9 & H & 3.357489 & 1.220761 & 0.027538 &
9 & H & 2.520166 & 1.721331 & 0.000001 &
9 & H & -2.634162 & 1.522759 & 0.000003 \\

10 & H & 3.287591 & -1.317508 & 0.089904 &
10 & H & 3.200139 & -0.742589 & -0.000001 &
10 & H & -3.014398 & -0.961092 & 0.000030 \\

11 & C & -2.350224 & 0.655942 & -0.033951 &
11 & C & -2.835888 & -0.424340 & 0.000000 &
11 & C & 2.582839 & -0.512889 & -0.000001 \\

12 & H & -2.880704 & 0.854119 & 1.781965 &
12 & H & -3.879451 & -0.136654 & 0.000000 &
12 & H & 3.613350 & -0.780863 & 0.000006 \\

13 & C & -1.156039 & 1.339324 & -0.093789 &
13 & C & -1.904100 & 0.518783 & 0.000001 &
13 & C & 1.930248 & 0.637439 & 0.000017 \\

14 & H & -1.089769 & 2.417337 & -0.145020 &
14 & H & -2.115003 & 1.579733 & 0.000002 &
14 & H & 2.274201 & 1.647357 & 0.000026 \\

15 & C & -2.336161 & -0.771455 & -0.073437 &\multicolumn{5}{c|}{}&\multicolumn{5}{c}{}\\

16 & H & -3.272839 & -1.314938 & -0.104679 &\multicolumn{5}{c|}{}&\multicolumn{5}{c}{}\\

17 & C & -1.145553 & -1.438097 & -0.048681 &\multicolumn{5}{c|}{}&\multicolumn{5}{c}{}\\

18 & H & -1.078372 & -2.518071 & -0.060908 &\multicolumn{5}{c|}{}&\multicolumn{5}{c}{}\\

\hline
\end{tabular}
}
\label{tab:ts3_p1_ts1}
\end{table*}
\FloatBarrier

\section{Relative energies of the isomers of the final product C$_8$H$_7$N$_2^+$}

\begin{figure*}[ht!]
    \centering
    \includegraphics[width=0.8\textwidth]{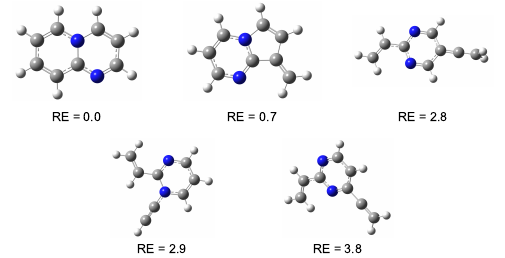}
    \caption{\textcolor{black}{Relative energies (RE in eV) of the lowest energy structures of the final product C$_8$H$_7$N$_2^+$ (m/z 131) calculated at the B3LYP/6-311G(d) level of theory.} 
    \label{isomer}}
\end{figure*}

\section{Intrinsic reaction coordinate (IRC) Plots}

IRC plots represent the energy variation along the reaction coordinate from the transition state to reactants and products. These plots help visualize the feasibility of the proposed reaction pathway.
\par Through TS-1, as mentioned in the main text, the hydrogen transfer happens from the pyrimidine carbon to the acetylene carbon. The IRC plot shown below, starting from this TS-1 is symmetrical, showing smooth transfer of H-atom, connecting the reactant and final product (P-1), the relative energy of which w.r.t TS-1 matches that mentioned in the reaction profile diagram (Figure~\ref{fig4} in the main text)
\par Through TS-2, the C-C bond formation takes place, leading to cyclisation. The IRC plot shown below, the product, smoothly connecting to the TS-2, after cyclisation, is with lower energy than the reactant energy, and the relative energy matches that reported in the energy profile diagram in the main text (Figure~\ref{fig4}).
\par Through TS-3, finally the H-atom elimination happens. However, the product energy (P-2) after deprotonation is higher than that of the reactant (INT-3). This also matches the relative energy trend in the energy profile diagram (Figure~\ref{fig4})

\begin{figure*}[ht!]
\centering

\begin{subfigure}{0.3\textwidth}
    \centering
    \includegraphics[width=\linewidth]{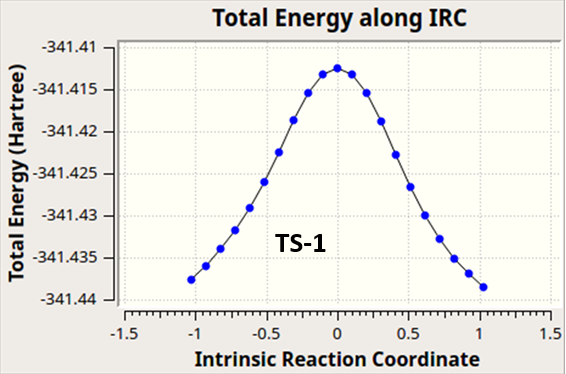}
    \caption{TS-1}
\end{subfigure}
\hfill
\begin{subfigure}{0.3\textwidth}
    \centering
    \includegraphics[width=\linewidth]{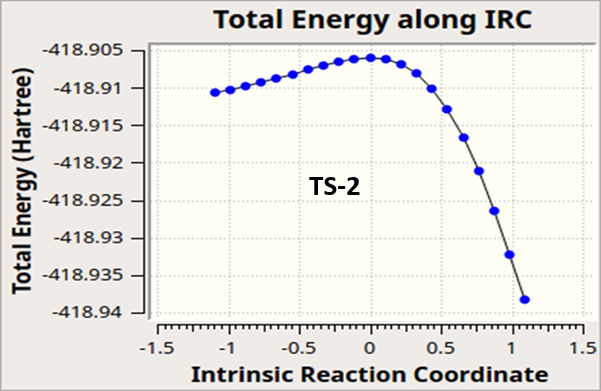}
    \caption{TS-2}
\end{subfigure}
\hfill
\begin{subfigure}{0.3\textwidth}
    \centering
    \includegraphics[width=\linewidth]{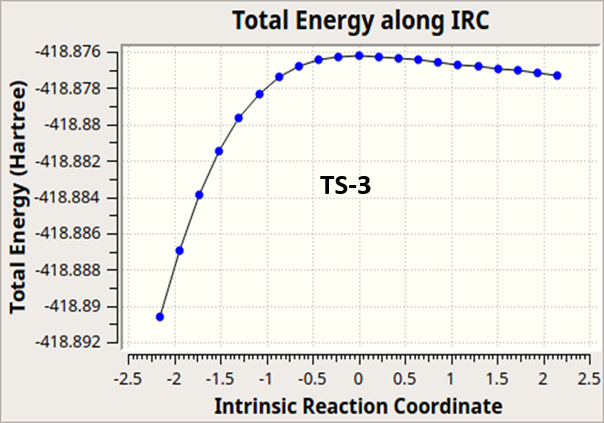}
    \caption{TS-3}
\end{subfigure}

\caption{IRC plots for TS-1, TS-2, and TS-3. The calculations confirm that each transition state smoothly connects the corresponding reactant and product minima on the potential energy surface.}
\label{fig:IRC}
\end{figure*}

\end{appendix}
\end{document}